%% file: gradient_pcap.tex
\documentclass[journal=jacsat,manuscript=article]{achemso}

\usepackage[version=3]{mhchem} 

\usepackage{graphicx}
\usepackage{dcolumn}
\usepackage{bm}

\usepackage[utf8]{inputenc}
\usepackage[T1]{fontenc}
\usepackage{mathptmx}
\usepackage{etoolbox}
\usepackage{xcolor}
\usepackage{amsmath}
\usepackage{booktabs}
\usepackage{longtable}
\usepackage{comment}
\usepackage{derivative}
\usepackage{multirow}
\usepackage{url}

\author{Soubhik Mondal}
\author{Ksenia B. Bravaya}
\email{bravaya@bu.edu}
\affiliation{Department of Chemistry, Boston University\\
    Boston, Massachusetts, 02215, USA}
	\title{Complex potential energy surfaces: gradients with projected CAP technique}

\usepackage{multirow}
\def\etal{\textit{et al.} }
\def\a0{$a_0$~}

\SectionNumbersOn
\renewcommand{\thesection}{\arabic{section}.}
\renewcommand{\thesubsection}{\thesection\arabic{subsection}.}
\renewcommand{\thesubsubsection}{\thesubsection\arabic{subsubsection}.}

\begin{document}
	
	\begin{abstract}
		The complex absorbing potential (CAP) technique is one of the commonly used Non-Hermitian quantum mechanics approaches for characterizing electronic resonances. CAP combined with various electronic structure methods has shown promising results in quantifying the energies and widths of electronic resonances in molecular systems. While CAP-based methods can be used to map complex potential energy surfaces for resonance states, efficient exploration of these surfaces, e.g. geometry optimization or dynamical simulations, require information on the nuclear gradient. Currently, the only nuclear gradients available for CAP-based methods are for Hartree-Fock and Equation-of-Motion Coupled-Cluster method with single and double excitations (J. Chem. Phys. 146, 031101 (2017)). Here we provide a general approach that relies on projected CAP formulation and extends gradients and non-adiabatic couplings formulations developed for bound-state electronic structure methods to resonances. The approach is not limited to a specific electron structure method and is generally applicable to any electronic structure methods, provided the information on the gradients and non-adiabatic couplings is available for bound states. Here, we focus on the State-Averaged Complete Active Space Self-Consistent Field (SA-CASSCF) and  Multi-Reference Configurational Interaction with Single excitation (MR-CIS) as our methods of choice.  We establish the accuracy of the developed gradients and report equilibrium geometries for several representative temporary anion species ($\mathrm{N_2^-}$, $\mathrm{H_2CO^-}$, $\mathrm{H_2CO_2^-}$ and $\mathrm{C_2H_4^-}$).
		
	\end{abstract}
	\maketitle
	
	\section{\label{sec:intro}Introduction}
	
Metastable electronic states are key precursors in reactive electron-molecule scattering processes. They play a key role in the breakage of single and double strands in DNA upon interaction with slow electrons\cite{boudaiffa2000resonant,alizadeh2015biomolecular,tonzani2006low,li2010low} and have also been proposed as intermediates for the formation of stable anions in interstellar medium\cite{mason2014electron,boyer2016role, petrie2007ions,millar2017negative}. These temporarily states, or resonances, lie in the electron detachment/ionization continuum, and their lifetimes vary from femtoseconds to picoseconds\cite{simons2008molecular, jordan1987temporary,taylor1966qualitative, simons1987ab}, which is long enough to be studied experimentally. Electronic resonances formed upon electron attachment can decay via several decay pathways accessible to the molecule. For short-lived anionic states, when the molecule does not leave the Franck-Condon region, an outgoing electron is emitted via autoionization and the molecule can end up in a vibrationally- excited state of the neutral, the process known as a resonance vibrational excitation (RVE)\cite{Allan_expt,itikawa_expt, VicicMar1996}. For longer-lived metastable states or steep dissociative surfaces for a metastable state, the molecule can undergo fragmentation, as is the case for dissociative electron attachments\cite{bald2008isolated,arumainayagam2010low,bass2003dissociative,schurmann2017resonant}.   Although single-point energy calculations can provide useful information on the energies and lifetimes of metastable states, a more thorough characterization of the processes and chemical reactions that proceed through electronic resonances requires tackling nuclear dynamics in metastable electronic states and disentangling nuclear motion and electronic decays, which often occur in comparable time-scapes\cite{santra2002non,santra2000interatomic,vib_fingerprint}.
    
As resonances belong to the continuous spectrum of the electronic Hamiltonian, they cannot be straightforwardly described with bound-state methods. Special techniques have to be used\cite{santra2002non, moiseyev1998quantum, domcke1991theory,langhoff1979stieltjes,mandelshtam1995spectral,hazi1970stabilization}. Moreover, a resonance is not represented by single electronic states in the Hermitian formulation\cite{jagau2017extending, jagau2016characterizing, jagau2022theory}. However, from the standpoint of nuclear dynamics simulations, it is convenient to operate with a single electronic state that represents a resonance both for model development and results interpretation. The solution is provided by Non-Hermitian Quantum Mechanics (NHQM) methods \cite{moiseyev2011non, moiseyev1998quantum, santra2002non}. In this case, the resonances are associated with a single square-integrable state of modified non-Hermitian electronic Hamiltonian with complex Siegert-Gamow eigenvalue\cite{gamow1928quantentheorie, siegert1939derivation}:
\begin{equation}
    E(\textbf{R})=E_R(\textbf{R})-i\frac{\Gamma(\textbf{R})}{2}
    \label{cpes}
\end{equation}	 
where $\mathrm{E_R}(\textbf{R})$ and $\Gamma(\textbf{R})$ are geometry-dependent ($\textbf{R}$) resonance energy and width, respectively, forming complex potential energy surfaces (CPES). 
    
Describing resonance with a single eigen-state is particularly attractive for mixed quantum/ classical dynamics models. In this case, chemical processes can be modeled by simulating nuclear motion classically\cite{gyamfi2022ab,kossoski2020nonadiabatic} and if nuclear quantum effects are ignored\cite{Moiseyev2017}, the force vector can be taken as negative of the gradient of the real part of the complex energy, $\mathrm{E_R}$(\textbf{R}). The irreversible electronic decay into the continuum governed by the resonance width, $\mathrm{\Gamma}$(\textbf{R}). In fact, the first attempts have been made to use NHQM electronic structure methods in on-the-fly BO or surface hopping dynamics on complex potential energy surfaces\cite{kossoski2020nonadiabatic, gyamfi2022ab}. Apart from evaluating accurate complex energies, these methods necessitate evaluation of forces on complex surfaces and till now, analytic gradients have only been available for Hartree Fock (CAP-HF) and equation of motion coupled-cluster method for electron attachment with single and double substitution (CAP-EOM-CCSD) levels of theory\cite{benda2017communication}.

Here we present a general approach for computing analytic nuclear gradients for CPESs that uses analytic gradients and non-adiabatic couplings for discretized continuum states obtained from Hermitian electronic structure calculations as input. The approach relies on the projected complex absorbing potential (pCAP) method\cite{ehara2012cap,gayvert2022projected,kanazawa2016low,thodika2022projected}, which extends hermitian quantum mechanics methods to the NH domain. The pCAP method has been   previously shown to produce accurate estimates of the resonance energies and widths for shape resonances in molecular systems\cite{ehara2012cap,kanazawa2016low,thodika2022projected,gayvert2022projected}. pCAP with careful choice of electronic structure methods has also been demonstrated to yield smooth CPESs\cite{jagau2014complex,gayvert2022projected,mondal2024complex}. Moreover, the quality of the pCAP CPESs was directly tested by comparing the computed cross-sections for resonant vibrational excitation (RVE) upon electron impact on N$_2$ with the experimental ones: the cross-sections simulated with the wave-packed dynamics on the pCAP CPESs are in very good agreement with the experiment\cite{mondal2024complex}. 

The proposed strategy can be implemented for any electronic structure method, for which nuclear gradients and non-adiabatic couplings are available. Here, for the proof-of-concept implementation and for testing the accuracy of the resulting nuclear gradients we focus on two electronic structure methods of moderate computational cost: state-averaged complete active-space self-consistent field (SA-CASSCF) and multi-reference configuration interaction with single excitations (MR-CIS). While these methods, especially SA-CASSCF, might not sufficiently account for electron correlation, and, thus, might not yield accurate complex potential surfaces, the presented results show promise for CPESs nuclear gradient evaluations with more advanced quantum chemistry methods, e.g.  multistate complete active space perturbation theory (MS-CASPT2)\cite{macleod2015communication,vlaisavljevich2016nuclear,park2017analytical,park2017fly}. 

The structure of the article is as follows. Sec. \ref{sec:thoery} introduces pCAP method (Sec.~\ref{subsec:Projected CAP}), describes the proposed model for analytic gradients evaluation (Sec.~\ref{sec:grad_eval}), and summarizes the state tracking approach employed for geometry optimizations (Sec.~\ref{section:tracking}). Sec. \ref{sec:comp_protocol} discusses system-specific details of the computational setup. Finally, computed analytic gradients are validated by comparison to the numerical ones (Sec. \ref{section:validation}) and by analyzing optimized geometries of representative metastable anions (Sec. \ref{section:opt}).

\section{\label{sec:thoery}Theory}
	\subsection{\label{subsec:Projected CAP} Projected CAP}
	CAP approach belongs to a group of NHQM methods, which describe an electronic resonance as a single eigen-state of a modified non-Hermitian Hamiltonian\cite{moiseyev2011non,moiseyev1998quantum,santra2002non}. In CAP method, the Hermitian electronic Hamiltonian ($\hat{H}$) is augmented by an imaginary absorbing potential ($-i\eta \hat{W}$)\cite{riss1993calculation}:
	\begin{equation}
		\hat{H}^\text{CAP}=\hat{H}-i\eta \hat{W}
		\label{capham}
	\end{equation}
where $\eta$ is the CAP strength parameter, and $\hat{W}$ is a one-particle operator specifying the functional form of CAP\cite{santra1999electronic,sommerfeld2015complex,ehara2016projected}. A resonance then appears as a single square-integrable eigen-state of $\hat{H}^\text{CAP}$ with a complex eigenvalue (Eq.~\ref{cpes}). Introduction of CAP absorbs the outgoing tail of the metastable wave function in the asymptotic limit and renders the wave-function square-integrable\cite{riss1993calculation}. The most common $\mathrm{\hat{W}}$  choices are the cuboid \cite{santra1999electronic}  and Voronoi CAP\cite{sommerfeld2015complex,ehara2016projected}. Here, we have used the cuboid CAP specified as follows:
\begin{equation}
	W = \sum_\alpha W_\alpha
\label{W_sum}
\end{equation}
\begin{equation}
	W_{\alpha} = \begin{cases}
		0 & \mathrm{, if } |r_{\alpha}-r^\mathrm{COM}_{\alpha}| \leq o_{\alpha} \\
		(\left | r_{\alpha}-r^\mathrm{COM}_{\alpha} \right| - o_{\alpha})^2 & \mathrm{, if } |r_{\alpha}-r^\mathrm{COM}_{\alpha}| > o_{\alpha}
	\end{cases}
\label{cuboid_cap}
\end{equation}
where $\alpha$ specifies the direction ($\alpha=x,y,z$),   $o_{\alpha}$ is the CAP onset in the corresponding direction, and $ r_{\alpha}$ is the value of the coordinate in a given direction. $r^\mathrm{COM}_{\alpha}$ is the center of mass (COM) coordinate in $\alpha$ direction for the molecular system defined as $r^\mathrm{COM}_{\alpha}=\frac{\sum_i m_iR_{i,\alpha}}{\sum_i m_i}$. $R_i$ and $m_i$ represents coordinate and mass of $i^\mathrm{th}$ nucleus respectively. We use a center of mass oriented grid instead of a center of nuclear charges one\cite{benda2017communication}, which is merely a choice of convenience. As follows from Eq.~\ref{capham}, the $\hat{H}^\text{CAP}$, and consequently its eigen-values, depends on the strength parameter $\eta$. In a complete basis set limit, the exact eigenvalues are recovered in the limit $\eta\rightarrow0$\cite{riss1993calculation}. For a finite basis, which is the case for practical calculations, one extracts the resonance position and width using an optimal value of $\eta$, $\eta_{opt}$, obtained by analyzing stationary points of $\eta$-trajectory, a series of complex energies computed for different values of $\eta$. In this work, the $\eta_{opt}$ was located using minimal logarithmic velocity criterion applied to uncorrected complex energies: $\frac{\eta dE(\eta)}{d\eta}\rightarrow \text{min} $\cite{riss1993calculation}. 

By using CAP-augment Hamiltonian, $\hat{H}^{CAP}$, instead of regular Hermitian Hamiltonian one can extend any quantum chemistry method originally formulated for bound electronic states to resonances. Indeed, CAP approach has been used together with multiple single- and multi- reference electronic structure methods, including symmetry adapted cluster–configuration inter-
action (SAC-CI)\cite{ehara2012cap}, multi-reference perturbation theory (MRPT2)\cite{kunitsa2017cap,phung2020combination}, equation of motion coupled-cluster (EOM-CC)\cite{bravaya2013complex,zuev2014complex,white2017second,ghosh2012equation}, algebraic diagrammatic construction (ADC)\cite{santra2002complex,feuerbacher2003complex}, and multi-reference configuration interaction (MR-CI)\cite{honigmann2006complex,honigmann2010use,sommerfeld2001efficient,sommerfeld1998temporary}. CAP approach can be integrated with an electronic structure method in two ways. First, one can operate with CAP-augmented Hamiltonian explicitly, which on practical level means that one has to modify the appropriate solvers and software to operate with complex Hamiltonians. Another consequence of explicit use of $\hat{H}^{CAP}$ is the necessity of performing multiple electronic structure calculations to characterize a single resonance's energy and width\cite{ghosh2012equation,jagau2014fresh,zuev2014complex,jagau2014complex}. This approach has been, for example, used to implement CAP-EOM-CCSD methods\cite{zuev2014complex}. An alternative solution is provided by the projected CAP scheme, originally proposed by Ehara and Sommerfeld\cite{sommerfeld2001efficient}. In this case, one starts with a regular Hermitian calculation employing Hermitian electronic Hamiltonian, $\hat{H}$ (Eq.~\ref{capham}), and generates a basis of discretized continuum states, or zero-order (ZO) basis ($\{\Phi^{ZO}_i\}$). ZO states are, therefore, eigen-states of the parent  Hermitian $\hat{H}$ approximated using a given model chemistry. The $\hat{H}^{CAP}$ is then projected on the ZO basis, resulting $\mathbf{H^{CAP}}$ matrix

		\begin{equation}
	\mathbf{H^{CAP}}=\mathbf{H^\text{ZO}}-i\eta \mathbf{W}
	\label{capham_proj}
\end{equation}	
,where $\mathbf{H^{ZO}}$ and $\mathbf{W}$ are matrices of electronic Hamiltonian and $\hat{W}$ in ZO basis. The former is most commonly diagonal (or can be made such, e.g. in the case of some flavors of multi-state perturbation theory).
$\mathbf{W}$ can be evaluated using one-particle reduced density and transition density matrices in atomic orbital (AO) basis ($\gamma^{AO}_{ij}$) for ZO states as follows:

\begin{equation}
\mathbf{W}_{ij}=Tr[\gamma_{ij}W^{AO}]
\label{eq:cap_m}
\end{equation}
,where $W^{AO}$ is CAP matrix calculated in AO basis.

Diagonalization of $\mathbf{H^{CAP}}$ yields resonance as an eigen-state and resonance energy and width as real and imaginary parts of the corresponding complex eigenvalue\cite{sommerfeld2001efficient,gayvert2022projected}. It has been shown that it is often enough to include only moderate number of states in ZO basis (10-30) to get a converged value of the resonance energy and width\cite{gayvert2022projected}. This approach has been used to implement CAP-based configuration interaction (CI)\cite{sommerfeld2001efficient}, SAC-CI\cite{ehara2012cap}, EOM-CCSD\cite{gayvert2022projected}, extended multiconfigurational quasidegenerate perturbation theory(XMCQDPT2)\cite{kunitsa2017cap}, extended multistate complete active space perturbation theory (XMS-CASPT2)\cite{phung2020combination} MR-CI\cite{thodika2022projected}, and ADC\cite{belogolova2021complex,dempwolff2021cap}.

\subsection{\label{sec:grad_eval} Analytic nuclear gradients for pCAP electronic structure methods}
The derivations below assume the Hermitian form of $\mathbf{H^{ZO}}$ Hamiltonian such that $\mathbf{H^{CAP}}$ has a complex-symmetric form. Yet, similar framework can be applied, for example, to non-Hermitian similarity-transformed Hamiltonian in EOM-CC methods by exploiting biorthogonal left and right eigen-states spaces. To set up the derivations, we first introduce a $\mathbf{H^{diag}}$ matrix, which is obtained by diagonalizing complex-symmetric $\mathbf{H^{CAP}}$ by a complex orthogonal transformation matrix $\mathbf{C}$, a matrix of $\mathbf{H^{CAP}}$  eigenvectors:

\begin{equation}
    \mathbf{
	H^{\text{diag}} =  C^T H^{\text{CAP}} C = C^T [{H^\text{ZO}}-i\eta \mathbf {W}]C}
	\label{diagon}
\end{equation}
One of the diagonal matrix elements of $\mathbf{H^{\text{diag}}}$ is, therefore, the complex eigenvalue associated with the resonances. Owing to the complex-symmetric structure of $\mathbf{H^{CAP}}$ left eigenvectors are transposed of the right ones, and, therefore, C matrix is orthogonal: $\mathbf{CC^T}=1$. Equation \eqref{diagon} specifies $\mathbf{H^{CAP}}$ transformation upon the basis change rotation from ZO to the diagonal basis, the basis of eigen-states of $\mathbf{H^{CAP}}$.
 
The nuclear gradient of the complex energy associated with the resonance can be, therefore, obtained from the gradient matrix written as follows: 
    
\begin{equation}
	\begin{aligned}		
        \nabla_R \mathbf{ H^{\text{diag}} }&= \nabla_R \mathbf{ \left[ C^T H^{\text{CAP}} C \right] }\\ 
		&=  \mathbf {C^T \nabla_R H^{\text{CAP}} C + \nabla_R C^T H^{\text{CAP}} C + C^T H^{\text{CAP}} \nabla_R C},
	\end{aligned}
	\label{grad_hcap_1}
\end{equation}

With the two latter terms canceling out (see Sec. 2 of Supporting Information), the final expression relating the gradient matrix in diagonal and ZO basis has the following form: 

\begin{equation}
	\begin{aligned}
		\mathbf{ \nabla_R H^{\text{diag}}} &= \mathbf{ C^T \nabla_R H^{\text{CAP}} C}
	\end{aligned}
	\label{grad_hcap_2}
\end{equation}

Using Eq.~\eqref{capham_proj} one can further rewrite the gradient matrix $\mathbf{\nabla_R H^{\text{diag}}}$, referred to from here on as $\mathbf{G^{\text{diag}}}$, by separating the terms originating from electronic Hamiltonian in ZO basis ($\hat{H}$) and from CAP:

\begin{equation}
	\begin{aligned}
	\mathbf{{G^{\text{diag}}} = {\nabla_R H^{\text{diag}}}} &= \mathbf{C^T \nabla_R \left [H^{\text{ZO}}-i\eta W \right ] C}\\
	&= \mathbf{C^T \nabla_R H^{\text{ZO}} C-i \eta C^T\nabla_R W C}\\
	&= \mathbf{C^T G^{\text{ZO}} C-i \eta U^T\nabla_R W C}
	\label{g_diag}
\end{aligned}	
\end{equation}
$\mathbf{G^{ZO}}$ in Eq.~\ref{g_diag} is the gradient matrix in ZO basis ($\mathbf{G^{ZO}}\equiv\nabla_R\mathbf{H^{ZO}}$), which has the following form: 
\begin{equation}
	\nabla_R \mathbf{H^{ZO}} = 
	\renewcommand{\arraystretch}{1.5}
	\begin{bmatrix}
		\langle \Phi^{ZO}_1 | \frac{\partial \hat{H}}{\partial R} | \Phi^{ZO}_1 \rangle & (E^{ZO}_2 - E^{ZO}_1) \langle \Phi_1^{ZO} | \frac{d\Phi_2^{ZO}}{dR} \rangle & \dots & (E^{ZO}_n - E^{ZO}_1) \langle \Phi_1^{ZO} | \frac{d\Phi_\mu^{ZO}}{dR} \rangle \\
		(E^{ZO}_1 - E^{ZO}_2) \langle \Phi_2^{ZO} | \frac{d\Phi_1^{ZO}}{dR} \rangle & \langle \Phi_2^{ZO} | \frac{\partial \hat{H}}{\partial R} | \Phi_2^{ZO} \rangle & \dots & (E^{ZO}_\mu - E^{ZO}_2) \langle \Phi_2^{ZO} | \frac{d\Phi_\mu^{ZO}}{dR} \rangle \\
		\vdots & \vdots & \ddots & \vdots \\
		(E^{ZO}_1 - E^{ZO}_\mu) \langle \Phi_\mu^{ZO} | \frac{d\Phi_1^{ZO}}{dR} \rangle & (E^{ZO}_2 - E^{ZO}_\mu) \langle \Phi_\mu^{ZO} | \frac{d\Phi_2^{ZO}}{dR} \rangle & \dots & \langle \Phi_\mu^{ZO} | \frac{\partial \hat{H}}{\partial R} | \Phi_\mu^{ZO} \rangle \\
	\end{bmatrix}
		\label{g_zo}
\end{equation}
where the diagonal elements and the off-diagonal elements are respectively the individual electronic state gradients and non-adiabatic couplings between the ZO states ($\{\Phi_{i}\}$), eigen-states of the Hermitian electronic Hamiltonian $\mathrm{\hat{H}}$. Importantly, all of the matrix elements of the gradient matrix in ZO basis, $\mathbf{G^{ZO}}$, are obtained from conventional Hermitian electronic structure calculations in the absence of CAP and can be computed by standard quantum chemistry packages without any code modification. 

The second term in the gradient matrix expression in diagonal basis, $\mathbf{G^{diag}}$ (Eq.~\ref{g_diag}), is the  CAP contribution, which can be calculated numerically using a stand-alone software.  A more detailed form of the Eq. \eqref{g_diag} yields the following final form of the $\xi$ state's gradient in the diagonal basis for n$^\text{th}$ atom in $\alpha$ direction:

\begin{equation}
\begin{aligned}
	{\mathbf{G^\text{diag}_{\xi\xi}}}\Big|_{n, \alpha} &= \sum_{\mu}\mathbf{C}^T_{\mu\xi}\mathbf{C}_{\mu\xi} \cdot \mathbf{G_{\mu\mu}^{ZO}}\Big|_{n, \alpha} + \sum_{\mu \neq \nu} \mathbf{C}^T_{\nu\xi}\mathbf{C}_{\mu\xi} \left [ E^{ZO}_{\mu}-E^{ZO}_{\nu} \right] \left\langle \Phi_{\nu}^{ZO}\Big | \pdv{\Phi_{\mu}^{ZO}}{R_{n, \alpha}}\right\rangle \\
	& - i\eta\cdot \sum_{\mu \nu} \mathbf{C}^T_{\nu\xi}\cdot{\pdv{W_{\mu\nu}}{R_{n, \alpha}}}  \cdot \mathbf{C}_{\mu\xi}
	\label{gdiag_main}
\end{aligned}	
\end{equation}
where $\mu$ and $\nu$ are correlated basis state indices.  

The ${\pdv{W_{\mu\nu}}{R_{n, \alpha}}}$ quantity from the last term in equation~\eqref{gdiag_main} can be expressed similarly to eqn.~\eqref{eq:cap_m} using the trace relation of one particle reduce density matrix in AO basis ($\mathrm{\gamma^{AO}}$). We further approximate that the $\mathrm{\gamma^{AO}}$ is impervious to change in nuclear coordinate $R_{n, \alpha}$, leading to,
\begin{equation}
\begin{aligned}
\pdv{W_{\mu\nu}}{R_{n, \alpha}} & =\pdv{}{R_{n, \alpha}} Tr\left[\gamma_{\mu\nu}^\mathrm{AO}W^\mathrm{AO} \right]\\
& \approx Tr\left[\gamma_{\mu\nu}^\mathrm{AO}\pdv {W^\mathrm{AO}}{R_{n, \alpha}} \right]
\end{aligned}
\label{capg_munu_trace}
\end{equation}

A closer inspection at the CAP gradient contribution $\pdv {W^\mathrm{AO}}{R_{n, \alpha}}$ in eqn.~\eqref{capg_munu_trace} reveals that it can be expanded into three contributions using eqn.~\eqref{W_sum}, 
\begin{equation}
\begin{aligned}
    \pdv {W^\mathrm{AO}}{R_{n, \alpha}} & =  \pdv {}{R_{n, \alpha}}\left\langle \chi_{\sigma} \left | W \right | \chi_{\rho} \right \rangle \\
     & =  \frac{m_n}{\sum_i m_i}\cdot \left\langle \chi_{\sigma} \left | \pdv{W_\alpha}{r^\mathrm{COM}_{\alpha}} \right | \chi_{\rho} \right \rangle + \left \langle \frac{\partial \chi_{\sigma}}{\partial R_{n,\alpha}} \left | \sum_\alpha W_\alpha \right | \chi_{\rho} \right \rangle + \left \langle \chi_\sigma \left | \sum_\alpha W_\alpha \right | \frac{\partial \chi_{\rho}}{\partial R_{n,\alpha}} \right \rangle ,
\end{aligned}
\label{capg_expansion}
\end{equation}
where $\rho$ and $\sigma$ are AO indices and $m_i$ is the atomic mass of $i^\mathrm{{th}}$ nucleus.

One arrives to the first term by taking derivative of W$_\alpha$ (presented in eqn.~\eqref{cuboid_cap}) with respect to the center of mass $r_\alpha^\mathrm{COM}$ movement in $\alpha$ direction. The analytical form of the $\pdv{W_\alpha}{r^\mathrm{COM}_{\alpha}}$ looks as follows, 
\begin{equation}
	\pdv{W_\alpha}{r^\mathrm{COM}_{\alpha}} = \begin{cases}
		0 & ,\text{if } |r_{\alpha}-r^\mathrm{COM}_{\alpha}| \leq o_{\alpha} \\
		-2\left ( r_{\alpha}- r^\mathrm{COM}_{\alpha}- o_{\alpha}\right) & ,\text{if } r_{\alpha}-r^\mathrm{COM}_{\alpha} > o_{\alpha}\\
		-2\left ( r_{\alpha}-r^\mathrm{COM}_{\alpha} +o_{\alpha}\right) & ,\text{if } r_{\alpha}-r^\mathrm{COM}_{\alpha} < -o_{\alpha}
	\end{cases}
	\label{CAPG_rg}
\end{equation}

In equation~\eqref{capg_expansion}, the last two parts correspond to contributions originating from the derivative of Gaussian basis function with respect to $R_{n, \alpha}$ similar to those derived in Ref~\citenum{benda2017communication}. Lastly, we keep the CAP onsets fixed in our geometry optimizations, hence, CAP derivative contributions with respect to $o_\alpha$ are ignored\cite{benda2017communication}.

Evidently, the gradients evaluated in eqn. \eqref{gdiag_main} are complex in nature. For geometry optimizations, the energy minimization is done for the real part of the CPES: $\mathrm{E_R}$ from \eqref{cpes}, hence only real part of the $\mathbf{G^{diag}}$ is extracted for the gradients of $\mathrm{E_R} (R)$.

\subsection{State tracking}
\label{section:tracking}
Geometry optimization using a method operating with multiple electronic states requires a reliable state tracking approach to avoid root switching. In this work we used two state tracking approaches. The first is based on evaluating of the wave-functions overlap for two consecutive steps. Second method uses attachment and detachment densities for state tracking\cite{head1995analysis,closser2014simulations}.

The overlap matrix is originally computed in the zero-order basis ($\mathbf{S^{ZO}}$). We used two approaches to calculate the overlap matrix. In first, the overlap was calculated by an external software WF-OVERLAP\cite{plasser2016efficient} using information on molecular orbitals and the ZO states' CI expansions as input. In the second, we used the built-in functionality of the OpenMolcas RASSI module\cite{MOLCAS8}.   The overlap matrix is then transformed to the diagonal basis ($\mathbf{S^{diag}}$) with the rotation matrices for the corresponding steps:

\begin{equation}
	\mathbf{S^\mathrm{diag}} =  \mathbf{C^T } \mathbf{ S^{ZO}} \mathbf{C'}
\end{equation} 

The optimization procedure follows the state that has the largest overlap with the state of interest from the previous step. 

For geometry optimization with symmetry on, ZO wave function overlaps are not available. In this case, attachment and detachment densities\cite{head1995analysis, closser2014simulations} are used for the state tracking as proposed by Closser \etal\cite{closser2014simulations}.

\section{\label{sec:comp_protocol}Computational details}

Geometry optimizations were carried out using geomeTRIC\cite{wang2016geometry} and optking\cite{heide2020optking} optimizers  with the energy and gradients provided to the optimizer on-the-fly. For the systems with the symmetry point group higher symmetry than $\mathrm{C_1}$, the optimization was performed with optking software\cite{heide2020optking} as symmetry is not implemented in geomeTRIC\cite{wang2016geometry}. CAP strength parameter $\eta$ was kept fixed during optimization at $\eta_{opt}$ value, optimal $\eta$ obtained for the equilibrium geometries of the neutral molecule. The box size $\mathrm{r^0_{\alpha=\left\{x,y,z\right\}}}$ was kept constant throughout the optimization. While this constraint can introduce some artifacts, it can be easily avoided by either using the Voronoi CAP\cite{sommerfeld2015complex,ehara2016projected} or adjusting the box CAP parameters along the optimization path. Exploring the effects of CAP onto the optimized equilibrium geometries of resonances is the subject of the future work. The complex energies evaluated with several considered model chemistries were computed with OpenCAP software interfaced\cite{gayvert} to OpenMolcas\cite{MOLCAS8} and COLUMBUS\cite{COLUMBUS} electronic structure packages. Zero-order gradients and non-adiabatic couplings for SA-CASSCF and MR-CIS calculations were computed with OpenMolcas\cite{MOLCAS8} and COLUMBUS\cite{COLUMBUS} packages, respectively. 

Our main choice of parent valence basis set is cc-pVTZ basis that has been further augmented by a subset of diffuse basis functions. Two types of diffuse subsets have been used. In the first, a single ghost atom (denoted as X) was placed in the center of mass of the molecule for each nuclear configuration (for every optimization step). In the second, the parent basis on the heavy atoms were augmented with several even-tempered diffuse basis functions.  The specific basis sets used for each of the considered electronic resonances are given in the corresponding section below.

The computational details specific to each studied system are summarized below.

\subsection{$\mathrm{N_2^-}$ anion: $^2\Pi_g$ resonance}
Equilibrium geometry of the $\mathrm{N_2^-}$ anion in the $^2\Pi_g$ resonance was obtained using two methods: State-averaged CASSCF (SA-CASSCF) and MR-CIS. The geometry optimization was performed within $\mathrm{D_{2h}}$ symmetry point group. State-averaging over 10 states was employed for SA-CASSCF: one $^1 A_g$ state representing the neutral and 9 anionic states of $^2 B_{2g}$ symmetry to represent the continuum. The active space chosen for this calculation was 5 electrons in 13 orbitals (1$b_{2u}$, 1$b_{3u}$, 5$b_{2g}$, 5$b_{3g}$, 1$b_{1u}$), where $\pi^*$ orbitals are represented by degenerate pairs of $b_{2g}$ and $b_{3g}$ orbitals.

MR-CIS calculations were carried out with smaller active spaces: (5e, 9o) and (7e,10o). Active space orbitals included for these two choices are (1$b_{2u}$, 1$b_{3u}$, 3$b_{2g}$, 3$b_{3g}$, 1$b_{1u}$) and (1$a_g$, 1$b_{2u}$, 1$b_{3u}$, 3$b_{2g}$, 3$b_{3g}$, 1$b_{1u}$) respectively. 
State-averaging over six states was used in both cases with one $^1A_g$ state and five anionic $^2 B_{2g}$ states representing the neutral and discretized continuum, respectively. 

MR-CIS calculations for the neutral and anionic states were performed using the same SA-CASSCF orbitals to get a balanced description of the neutral and anionic resonance states. No orbitals were frozen in the calculations\cite{mondal2024complex}. The starting N-N bond length for these three geometry optimizations is taken to be 1.095 \AA\ , corresponding to CASSCF(2e,2o) optimization of the neutral molecule with cc-pVTZ basis set.

To compare the equilibrium bond length and resonance energetic parameters with the previously reported data\cite{thodika2022projected,mondal2024complex} and precomputed MR-CIS curves\cite{thodika2022projected}, we also performed a SA-5-MR-CIS calculation with (5e, 8o) active space using the recipe prescribed by Thodika and Matsika\cite{thodika2022projected}. All the electronic structure calculations, including ZO energies\cite{szalay2012multiconfiguration}, densities, gradients and non-adiabatic couplings\cite{lischka2004analytic,dallos2004analytic} evaluations were performed with COLUMBUS electronic structure package\cite{COLUMBUS}.

\subsection{$\mathrm{H_2CO^-}$ anion: $\pi^*$ resonance}
SA-CASSCF geometry optimization of formaldehyde anion was performed with no symmetry constraints. Several state-averaging schemes have been used. In all cases, the state-averaging was performed over anionic states. The converged SA-CASSCF orbital were then used to perform CASCI calculation for the ground state of the neutral to get the reference energies. This choice of state-averaging scheme can introduce imbalance in description of the neutral and anion, which in turn can affect the computed resonance position. Yet, one can expect that not including the neutral in the state-averaging does not lead to detrimental description of the complex potential energy surface of the anionic resonance, and therefore, it's equilibrium geometry. As the primary focus of this work is on nuclear gradients for the complex potential energy surfaces, this choice of methodology does not affect the main conclusions of the work.  Two diffuse basis sets centered on the ghost atoms were used. The first basis set consisted of four $p$-type basis functions (4p). The second basis included two $s$-, five $p$- and two $d$- type even-tempered basis functions. In both cases, the initial exponents were chosen based on the most diffuse exponent in the parent basis. 
{\color{black}The scaled exponents are available in the Supporting Information}

MR-CIS geometry optimization was carried out using the $C_s$ symmetry point group. (3e, 8o) CASSCF active space consisted of 1 $a'$ ($\pi$) and 7$a''$($\pi^*$) orbitals with 2 electrons in $A'$ and an extra electron in $A''$ ($\mathrm{\pi^*}$). The state averaging over 7 $A''$($\pi^*$) anionic states was used. These CASSCF orbitals were used to perform subsequent anionic($^2 A''$) and neutral ($^1 A'$) MR-CIS calculations.

\subsection{Formic acid anion: $\pi^*$ resonance}

{\color{black} Calculations for formic acid anion were performed with SA-CASSCF and MR-CIS}. To assess the the basis set effects on the optimized geometry several different diffuse basis subsets were used: 3p, 4p, 5p, and 2s5p2d bases for SA-CASSCF.  The exponents are available in the Supporting Information.

{\color{black} For MR-CIS calculation, cc-pVDZ and cc-pVTZ basis sets were used with a ghost atom in the center of mass carrying [3p] and [4p] diffuse orbitals. (3e, 12o) active space was used for both basis sets. State-averaging over seven and eight states were used for cc-pVTZ+[3p] and cc-pVDZ+[4p] respectively.}

To explore the effects of interaction with the continuum on the geometry of the $\pi^*$ anion, we have also performed a reference geometry optimization for valence localized $\mathrm{\pi^*}$ state of formic acid, which was achieved by using the basis without any diffuse basis functions. The active space chosen for the latter calculation was 3 electrons over 4 orbitals and the valence $\mathrm{\pi^*}$ root of SA-CASSCF and subsequent MR-CIS were optimized.

\subsection{$\mathrm{C_2H_4^-}$ anion: $\pi^*$ resonance}
CASSCF level geometry optimization was carried out with no symmetry constraints. cc-pVTZ basis augmented with three extra diffuse $p$-type basis functions placed on the carbon atoms was used. A preliminary CASSCF/cc-pVTZ optimized geometry of the neutral was chosen as a starting point for the geometry optimization of the anion. 

MR-CIS geometry optimization was performed using the $\mathrm{C_{2h}}$ symmetry point group. Same basis set as discussed above was employed. The starting geometry was the same as for the SA-CASSCF geometry optimization. (3e, 10o) active space included one $\mathrm{b_u}$ and nine $\mathrm{a_g}$ orbitals. State-averaging over eight states of $\mathrm{A_g}$ symmetry was used. The reference MR-CIS energy of the neutral  was evaluated using the converged CASSCF orbitals.  

A smaller cc-pVDZ basis with three extra diffuse $p$-type orbitals placed on the carbon atoms was also used to assess the basis set effect on the equilibrium geometry of the anion.  The  (3e,7o) SA-CASSCF active space in this case included 1 $\mathrm{b_u}$ and 6 $\mathrm{a_g}$ orbitals. State averaging over six states of $\mathrm{A_g}$ symmetry was used. The same approach to that used for a larger basis was taken to evaluate the MR-CIS reference energy of the neutral. 

Similarly to formic acid anion, to explore the effects of the interaction with the continuum on the equilibrium geometry of the $\pi^*$ resonance of C$_2$H$_4^-$, CASSCF(2e,2o) geometry optimization for the valence localized anion was performed using cc-pVTZ basis.

\section{\label{sec:results}Results}

Below we discuss the performance and accuracy of the proposed form of the analytic gradients for the complex potential energy surfaces for pCAP-based methods. We first focus on comparison between analytic and numerical gradients. We then discuss the optimized geometries of representative metastable anions.

\subsection{Validation against the numerical gradients}
\label{section:validation}
To test the accuracy of the proposed analytic gradients, we have performed the comparison between analytic and numerical gradients. Note that, as approximations have been introduced, \textit{i.e.} {\color{black}the variation in the rotation matrix with the change in the nuclear coordinates for evaluating the CAP contribution, 
we do not anticipate the exact agreement. For numerical calculation of the gradient, 
we choose a $\Delta$R =0.0001\AA\ step in each direction for all the atoms and calculate the gradient using two-point finite difference. Fig.~\ref{fig:numerical_validation} shows the computed numerical and analytic gradients for MR-CIS (N$_2^-$) and SA-CASSCF (formic acid anion). We see a good agreement between the two sets of data, which points to a negligible contributions of the approximations made. Moreover, comparing the gradient terms originating from CAP for CAP-EOM-EA-CCSD\cite{benda2017communication} and pCAP-EOM-EA-CCSD points to only minor differences, which also supports the proposed form of the gradients (see Supporting Information).

\begin{figure}[!tbh]
	\centering
	\includegraphics[width=1.0\textwidth]{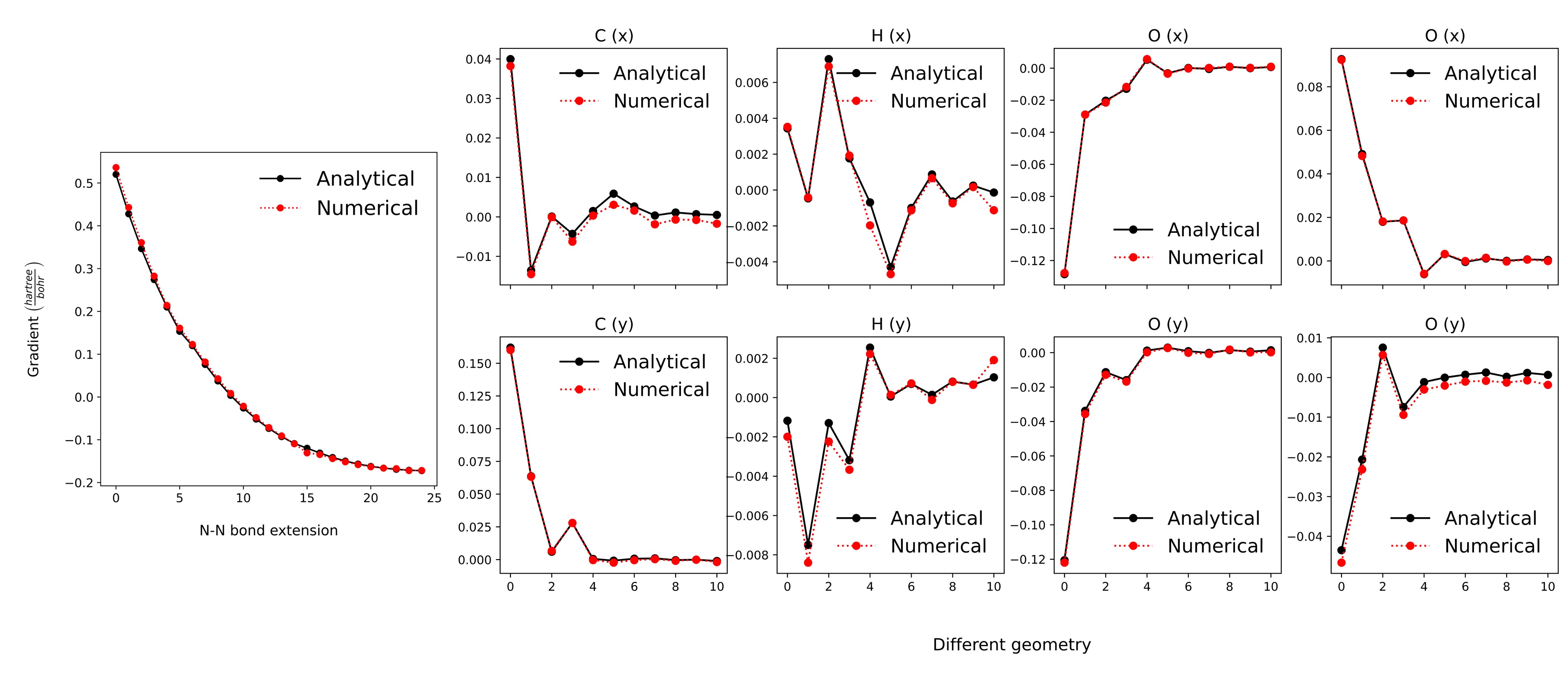}
	\caption{Numerical validation of the proposed analytic gradients: analytic vs. numerical gradients. Left panel shows the numerical and analytical gradients for {\color{black} one of} the nitrogen atoms' displacement in Z direction for the $^2\pi_g$ resonance state of $\mathrm{N_2^-}$ obtained for pCAP used with MR-CIS/SA-6-CASSCF(7e, 10o) and cc-pVTZ+[2s5p2d] basis set for different bond lengths. Right panel shows representative values of the gradients for different atoms in $\mathrm{H_2CO_2^-}$ for different geometries for pCAP-based SA-15-CASSCF(3e, 23o) method with cc-pVTZ+X[5p] basis. The top and bottom rows represent X and Y directions. The gradients along other displacements have much smaller values ($10^{-6}-10^{-5}$ hartree/bohr).}
	\label{fig:numerical_validation}
\end{figure}

\subsection{Equilibrium geometries of metastable anions}
\label{section:opt}
We have considered four representative shape resonances in molecular systems to explore the performance of the pCAP methods and the corresponding analytic gradients for predicting equilibrium geometries of metastable anions. Specifically, we have considered the shape resonances in the anions of dinitrogen ($\mathrm{N_2}$), formaldehyde ($\mathrm{H_2CO}$), formic acid ($\mathrm{H_2CO_2}$) and ethene ($\mathrm{C_2H_4}$). The resulting equilibrium geometries and relevant natural orbitals are shown in Figure~\ref{fig:optimized_structure}. Below we discuss each of the resonances separately.

\begin{figure}[!tbh]
    \centering
    \includegraphics[width=1.0\textwidth]{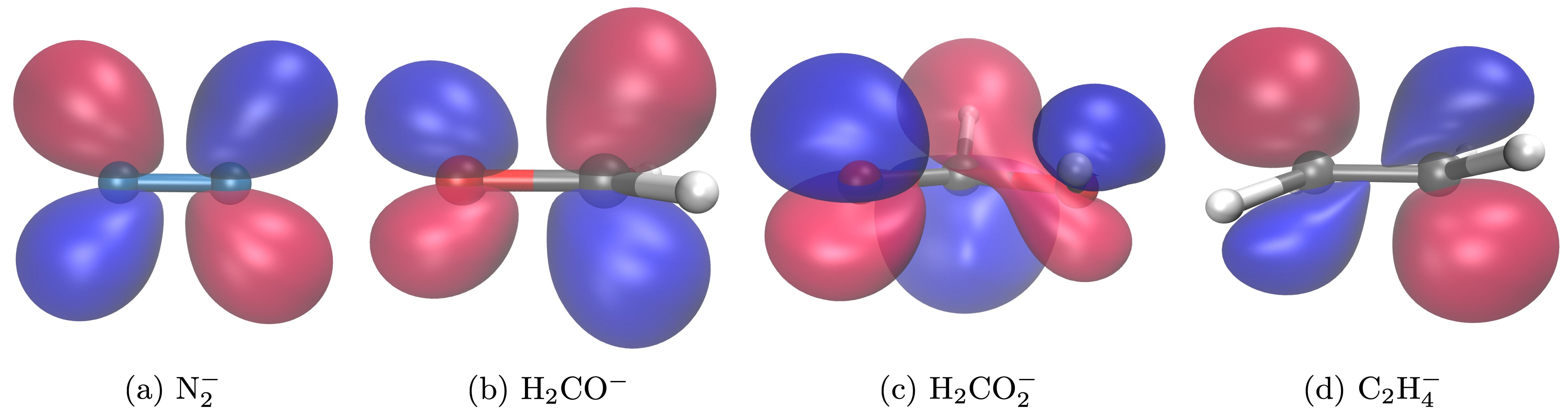}
    \captionsetup{singlelinecheck=off} 
    \caption{Natural orbitals (real part of natural orbitals after diagonalizing complex densities, see eq.(1) in Supporting Information) of anionic resonances of $\mathrm{N_2}$\textsuperscript{\emph{a}}, $\mathrm{H_2CO}$\textsuperscript{\emph{b}}, $\mathrm{H_2CO_2}$\textsuperscript{\emph{c}}, and $\mathrm{C_2H_4}$\textsuperscript{\emph{d}}. The $\mathrm{\pi^*}$ orbitals are plotted at the equilibrium structures of the respective resonant anions.}
    \footnotesize
    \textsuperscript{\emph{a}}MR-CIS/SA-6-CASSCF(7e,10o), cc-pVTZ+[2s5p2d] (D$_\mathrm{2h}$ symmetry);
    \textsuperscript{\emph{b}}MR-CIS/SA-7-CASSCF(3e,8o)/cc-pVDZ+[3p] (C$_\mathrm{s}$ symmetry);
    \textsuperscript{\emph{c}}SA-15-CASSCF(3e,23o)/cc-pVTZ+X[5p] (C$_\mathrm{1}$ symmetry);
    \textsuperscript{\emph{d}}MR-CIS/SA-6-CASSCF(3e,7o)/cc-pVTZ+[3p] ($\mathrm{C_{2h}}$ symmetry).
    \label{fig:optimized_structure}
\end{figure}
 
\subsubsection{$\mathrm{N_2^-}$ $^2 \Pi _g$ shape resonance}

Optimized bond lengths and energetic parameters for the  equilibrium geometries of the neutral and of the $\mathrm{N_2^-}$ $^2 \Pi _g$ shape resonance are listed in Table~\ref{tab:N2_opt}. Population of the $\pi^*$ orbital leads to the effective bond order decrease and bond elongation. Indeed, all method and basis set combinations correctly capture this behavior: the bond length increases by 0.06 - 0.10 \AA\ depending on the method. As mentioned earlier, the lack of dynamical electron correlation in SA-10-CASSCF(5e, 13o) leads to inaccurate description of the resonance energetic parameters.  The resonance energy of 4.4 eV at the equilibrium geometry of the neutral is more than 1 eV away from the experimental value\cite{berman1983nuclear} and from more accurate theoretical estimates\cite{sommerfeld2001efficient,jagau2014fresh,gayvert2022projected,thodika2022projected}. In addition, SA-10-CASSCF(5e, 13o) significantly overestimates the width of the resonance (Table~\ref{tab:N2_opt}). Taking into account dynamic electron correlation brings the resonance down in energy. MR-CIS/SA-6-CASSCF(5e, 9o) yields resonance energy and width at the equilibrium of the neutral of 3.2 and 0.54 eV, respectively, which are much closer to the experimental estimate\cite{berman1983nuclear}.  The equilibrium bond length of the neutral is identical to CASSCF results. The bond length for the optimized geometry of the anion is slightly greater than the corresponding SA-10-CASSCF(5e, 13o) value (by $\sim$0.02\AA ). We have also explored the effects of extending the active space by including additional $a_g$ occupied molecular orbital in the active space yielding MR-CIS/SA-6-(7e, 10o) model. The resulting changes in resonance energy, width, and equilibrium bond length are minor (Table~\ref{tab:N2_opt}). To compare the optimized geometries to previously reported data obtained with similar methods, we have performed MR-CIS/SA-10-CASSCF(5e,8o) calculations. One of the difference in the data reported in Ref.\citenum{thodika2022projected} and here is the number of ZO states used for pCAP: five states used here vs. ten states in Ref.~\citenum{thodika2022projected}. Thodika and Matsika have reported a shallow potential energy minimum close to 1.15\AA\ by scanning along the N--N distance with 0.05\AA\ step. A cubic interpolation of their potential energy curve data yields a minima at 1.17\AA\ . Using a similar MR-CIS/SA-10-CASSCF(5e, 8o) approach, we locate the minimum at 1.18\AA\ . The energies and the widths are very close to those reported in Ref.~\citenum{thodika2022projected} for both equilibrium geometries of the neutral and of the anion. 


Geometry relaxation in the anionic state brings the two states closer in energy by 0.5 - 1 eV depending on the methods, which is also reflected in the consistent decrease in the resonance width (Table~\ref{tab:N2_opt}). 

\begin{table}[!tbh]
\caption{Equilibrium bond length ($\mathrm{R_{N-N}}$, \AA), resonance energy ($\mathrm{E_R}$, eV), and resonance width ($\mathrm{\Gamma}$, eV) for the optimized geometries of the neutral ($\mathrm{N_2}$) and anion ($\mathrm{N_2^-}$) computed with different methods and basis sets.		
\protect{\label{tab:N2_opt}}}
\begin{center}
    \setlength{\tabcolsep}{4pt} 
        \renewcommand{\arraystretch}{1.2} 
		\begin{tabular}{lcccccc}
            \hline
			\multirow{2}{*}{\textbf{Methods/basis}} & \multicolumn{3}{c}{ \textbf{N$_2$}} & \multicolumn{3}{c}{\textbf{N$_2^-$}} \\ 
			&  $\mathbf{R_{N-N}}$& \textbf{E ($\mathbf{R_{N-N}}$)}  & \textbf{$\boldsymbol{\Gamma}$ ($\mathbf{R_{N-N}}$)} &  $\mathbf{R_{N-N}}$ & \textbf{E }($\mathbf{R_{N-N}}$) & $\boldsymbol{\Gamma}$ ($\mathbf{R_{N-N}}$) \\ 
			\hline
		SA-10-CASSCF(5e, 13o) &1.095 & 4.448 & 1.622 &1.155 & 3.982 & 1.070\\
			cc-pVTZ+[2s5p2d] & & & & & &\\ 
		MR-CIS/SA-6-CASSCF(5e, 9o) &1.095 & 3.166 &  0.544 &1.176 & 2.418& 0.346\\ 
			cc-pVTZ+[2s5p2d]& & & & & &\\ 
		MR-CIS/SA-6-CASSCF(7e, 10o) &1.095 & 3.516& 0.631&1.193 & 2.538 & 0.376\\ 
			cc-pVTZ+[2s5p2d]& & & & & &\\ 
		MR-CIS\textsuperscript{\emph{a}}/SA-10-CASSCF(5e,8o) &1.100\textsuperscript{\emph{b}} & 3.033 &  0.503 &1.180 & 2.226 & 0.329\\ 
		aug-cc-pVTZ+X[3s3p3d] & & & & & &\\ 
            \hline
		\end{tabular}
          \textsuperscript{\emph{a}} Five states have been used for pCAP calculations instead of ten states in Ref.~\citenum{thodika2022projected};
           \textsuperscript{\emph{b}} The bond length corresponds to B3LYP/aug-cc-pVTZ optimized geometry from Ref.~\citenum{thodika2022projected}. The geometry is used here for consistency of comparison with the data by Thodika et al.~\cite{thodika2022projected}.

        \end{center}   
\end{table}

\subsubsection{Formaldehyde anion: $\pi^*$ shape resonance}
Table.~\ref{tab:H2CO_opt} lists the geometric parameters for the optimized geometries of the formaldehyde in the ground state of the neutral and in the shape $\pi^*$ resonance state. As anticipated, populating the $\pi^*$ orbital leads to elongation of the C=O double bond: the bond length increases by 0.09-0.10\AA\ depending on the electronic structure method. The $\angle$H-C-O angles and C--H bond lengths do not change considerably. Overall, the geometries obtained with both SA-CASSCF and MR-CIS are quite similar.

\begin{table}[!tbh]
\caption{Geometrical parameters for CH$_2$O and CH$_2$O$^-$ in the corresponding equilibrium geometries computed using different electronic structure methods. Bond lengths and angles are reported in \AA~ and degrees, respectively.}
		\label{tab:H2CO_opt}
	\centering
    		\setlength{\tabcolsep}{2pt} 
                \renewcommand{\arraystretch}{1.2} 

		\begin{tabular}{lcccccc}
        \hline
			\multirow{2}{*}{\textbf{Methods/basis}} & \multicolumn{3}{c}{ \textbf{CH$_2$O}} & \multicolumn{3}{c}{\textbf{CH$_2$O$^-$}} \\ 
			& \textbf{R(C=O)} & \textbf{R(C–H)} & \textbf{$\angle$(H–C–O)} &  \textbf{R(C=O)} & \textbf{R(C–H)} & \textbf{$\angle$(H–C–O) }\\ 
            \hline
            SA-15-CASSCF(3e, 16o)/ & 1.178 & 1.093 & 122 & 1.278 & 1.110 & 120\\
			cc-pVTZ+X[4p] & & & & & & \\ 
			SA-19-CASSCF(3e, 20o)/ & 1.178 & 1.093 & 122 & 1.268 & 1.098 & 121 \\
			cc-pVTZ+X[2s5p2d] & & & & & & \\ 
			MRCIS/SA-7-CASSCF(3e, 8o)/ & 1.182 &1.101 & 122 & 1.282 & 1.102 & 122 \\
			cc-pVDZ+[3p]  (C$_s$ symmetry) & & & & & & \\ 
        \hline
		\end{tabular}
\end{table}

The resonance parameters for the equilibrium geometry of the neutral and of the $\pi^*$ shape resonance are listed in the Table~\ref{tab:H2CO_eta_params}. Similarly to what has been observed for the N$_2$, SA-CASSCF significantly overestimates resonance energy at the equilibrium geometry of the neutral placing the state at 2.4- 2.6 eV above the neutral.   Including of the dynamic electron correlation effects brings the resonance energy down to 0.9 eV, which is in much better agreement with earlier more accurate theoretical estimates\cite{ehara2012cap,choi1989electron,benda2017communication,kunitsa2019feshbach} and with the experimental reference value of 0.86\cite{burrow1976electron}. Interestingly, the widths obtained with the three methods rather close (variation within 0.2 eV). Notably, while extension of the diffuse basis set from [4p] to [2s5p2d]  leads to the shift in resonance energy of 0.2 eV and to the increase in the width of 0.08 eV, the effects on the optimized geometry are negligible (Table~\ref{tab:H2CO_opt}). Same trends for resonance energetic parameters, i.e. SA-CASSCF overestimating the resonance position and MR-CIS bringing the energy down, are observed for other molecular shape resonances, and, therefore, below we only focus on the equilibrium geometries rather than energetic parameters, which are reported in Supporting Information.

\begin{table}[!tbh]
\caption{Energy ($E_R$, eV) and width ($\Gamma$, eV) of the $\pi^*$ resonance in the equilibrium geometries of the neutral (CH$_2$O) and of the anion (CH$_2$O$^-$) computed with different electronic structure methods.}
	\label{tab:H2CO_eta_params}
	\centering
        \setlength{\tabcolsep}{8pt} 
        \renewcommand{\arraystretch}{1.2} 
		\begin{tabular}{lcccc}
            \hline
			\textbf{Methods/basis} & \multicolumn{2}{c}{ {$\mathbf{H_2CO}$}} & \multicolumn{2}{c}{ {$\mathbf{H_2CO^-}$}} \\ 
			& $\mathbf{E_R}$ & $\boldsymbol{\Gamma}$ & $\mathbf{E_{R}}$ & $\boldsymbol{\Gamma}$ \\ \hline
			SA-15-CASSCF(3e, 16o)/ & 2.426 & 0.338  & 1.281 & 0.232 \\
			cc-pVTZ+X[4p] & & & & \\ 
			SA-19-CASSCF(3e, 20o)/ & 2.633 & 0.416 &  1.086 & 0.687 \\
			cc-pVTZ+X[2s5p2d] & & & & \\ 
			MRCIS/SA-7-CASSCF(3e, 8o)/ & 0.931 & 0.211 & 0.168 & 0.101 \\
			cc-pVDZ+[3p]  & & & & \\ 
            \hline
        \end{tabular}
\end{table}

\subsubsection{Formic acid anion: $\pi^*$ shape resonance}
Formic acid undergoes significant structural changes in the $\pi^*$ resonance relative to the neutral\cite{benda2017communication,benda2018structure}. In addition to the single C--O, double C=O bonds elongation geometry relaxation is accompanied by the out-of-plane distortion\cite{benda2017communication,benda2018structure}.  Table~\ref{tab:H2CO2_opt} reports the equilibrium values of the relevant structural parameters for the neutral and anionic $\pi^*$ states of formic acid.  

All methods consistently reproduce the increase in the C=O and C--O bond lengths of 0.08 - 0.09\AA\ and 0.10 - 0.14\AA\, respectively. Most variation between the methods is observed in the predicted magnitude of the $\langle$H-C-O-H dihedral angle. pCAP-based SA-CASSCF methods yield the angle of 116-118 $^\circ$, while pCAP-based MR-CIS yields the value of 134$^\circ$. Owing to these notable structural changes, we have chosen formic acid anion as a model system to explore the effects of the size of the diffuse basis subset on the resulting structural parameters. As follows from Table~\ref{tab:H2CO2_opt} addition of extra diffuse basis functions beyond 3$p$-type functions on the ghost atom does not lead to significant changes in the optimized geometry configurations of the anion. Importantly, the SA-3-CASSCF(3e, 4o) geometry optimization approximating the resonance by a localized valence $\pi^*$ state ({\color{black}penultimate row} in the Table~\ref{tab:H2CO2_opt}) yields bond lengths and the dihedral angle that are rather different from the comparable pCAP SA-CASSCF calculations. The localized simulation overestimated the deviation from the planarity with the resulting $\langle$H-C-O-H 134$^{\circ}$ in comparison to 116-118$^{\circ}$ for pCAP-based calculations. Similar trend is observed for MR-CIS method, where geometry optimization of a localized $\pi^*$ resonance yields a structure with a stronger out-of-plane relative to the pCAP-based calculation: 140$^{\circ}$ vs 134$^{\circ}$, respectively. These discrepancies in the optimized structural parameters of the resonance and approximate localized state point to the necessity of the proper account of the metastable character of the state.

These set of observations establish several importance of pCAP method based optimizations. Firstly, representing resonance state by bound state calculation does not provide the same description of the resonance. The geometrical parameters compared in the Table ~\ref{tab:H2CO2_opt} shows that even though it may be possible to represent the resonance state by valence bound state, it does not necessarily guarantee valid results. Secondly, the convergence with respect to the basis sets size provides the glimpse that just adding 3 diffuse orbitals in the ghost atom is reasonably good enough to describe the the discretized continuum.
\begin{table}[!ht]
\caption{Representative optimized structural parameters of HCOOH and HCOOH$^-$ computed with different methods. Bond lengths and angles are reported in \AA~ and degrees, respectively.}
	\label{tab:H2CO2_opt}
	\centering
    		\renewcommand{\arraystretch}{1.2} 
    		\setlength{\tabcolsep}{0.6pt} 
		\begin{tabular}{lcccccc}
        \hline
			\multirow{2}{*}{\textbf{Methods/basis}} & \multicolumn{3}{c}{ \textbf{HCOOH}} & \multicolumn{3}{c}{\textbf{HCOOH$^-$}} \\ 
			& \textbf{R(C=O)} & \textbf{R(C–OH)} & \textbf{$\angle$(H–C–O-H)} & \textbf{R(C=O)} & \textbf{R(C–OH)} & \textbf{$\angle$(H–O-C–H)} \\ 
			\midrule
			SA-16-CASSCF(3e, 17o)/ & 1.178 & 1.319 & 180 & 1.264 & 1.434 & 117 \\
			cc-pVTZ+X[3p] & & & & & & \\ 
			SA-15-CASSCF(3e, 23o)/ & 1.178 & 1.319 & 180 & 1.261 & 1.428 & 118 \\
			cc-pVTZ+X[4p] & & & & & & \\ 
			SA-15-CASSCF(3e, 23o)/ & 1.178 & 1.319 & 180 & 1.267 & 1.426 & 118 \\
			cc-pVTZ+X[5p] & & & & & & \\ 
            SA-16-CASSCF(3e, 17o)/ & 1.178 & 1.319 & 180 & 1.265 & 1.430 & 116 \\
			cc-pVTZ+X[2s5p2d] & & & & & & \\ 
			MRCIS/SA-7-CASSCF(3e, 12o)/ & 1.178 & 1.319 & 180 & 1.269 & 1.439 & 132 \\
			cc-pVTZ+X[3p] & & & & & & \\ 
			MRCIS/SA-8-CASSCF(3e, 12o)/ & 1.180 & 1.322 & 180 & 1.267 & 1.465 & 134 \\
			cc-pVDZ+X[4p] & & & & & & \\ 
			SA-3-CASSCF(3e, 4o)/ & 1.178 & 1.319 & 180 & 1.248 & 1.420 & 134 \\
			cc-pVTZ (valence-$\pi^*$, w/o CAP) & & & & & & \\ 
			MR-CIS/SA-3-CASSCF(3e, 4o)/ & 1.178 & 1.319 & 180 & 1.265 & 1.462 & 140 \\
			cc-pVTZ (valence-$\pi^*$, w/o CAP) & & & & & & \\ 
            \hline
		\end{tabular}
\end{table}


\subsubsection{$\mathrm{C_2H_4^-}$: $\pi^*$ shape resonance}
Finally, we present the results of the geometry optimization of the ethylene anion in the $\pi^*$ resonance state. SA-CASSCF as well as the two MR-CIS calculations yield very similar results. Both sets of calculations reproduce bond extension in the $\pi^*$ state of 0.08-0.10 \AA. The dihedral angle is slightly higher for SA-15-CASSCF  (37$^{\circ}$) in comparison to the MR-CIS (30$^{\circ}$). of the Geometry optimization with and without symmetry has almost similar geometry change for ethylene $\pi^*$ temporary anion. The distortion from planarity is also observed and is even more pronounced in the CAP-free calculations (42$^\circ$). Benda and Jagau\cite{benda2017communication} has reported a dihedral angle change of $\sim$27$^\circ$ for a CAP-EOM-EA-CCSD based optimization with aug-cc-pVDZ+3s3p(A) basis, which is quite close to MR-CIS results reported here.  

\begin{table}[!tbh]
\caption{Geometrical parameters of $\mathrm{C_2H_4}$ and $\mathrm{C_2H_4^-}$ obtained with different methods. Bond lengths and angles are reported in \AA~ and degrees, repsectively.}
	\label{tab:C2H4_opt}
	\centering
		\setlength{\tabcolsep}{9pt} 
    	\renewcommand{\arraystretch}{1.2} 
		\begin{tabular}{lcccc}
        \hline
        \multirow{2}{*}{\textbf{Methods/basis}} & \multicolumn{2}{c}{ $\mathbf{C_2H_4}$} & \multicolumn{2}{c}{\textbf{$\mathbf{C_2H_4}^-$}} \\ 
			& \textbf{R(C=C)} & \textbf{$\angle$(H–C–C-H)} &  \textbf{R(C=C)} & \textbf{$\angle$(H–C–C-H)} \\ \hline
            SA-15-CASSCF(3e, 20o)/ & 1.330 &0 & 1.429 & 37 \\
			cc-pVTZ+[3p] & & & &\\ 
			MR-CIS/SA-8-CASSCF(3e, 9o)/ & 1.341 & 0 & 1.421 & 30 \\
			cc-pVTZ+[3p] ($\mathrm{C_{2h}}$ symmetry) & & & &\\ 
			MR-CIS/SA-6-CASSCF(3e, 7o)/ & 1.341 & 0	 & 1.426 & 31 \\
			cc-pVDZ+[3p] ($\mathrm{C_{2h}}$ symmetry) & & & &\\ 
			CASSCF(3e, 2o)/ & 1.330 & 0 & 1.437 & 42\\
			cc-pVTZ (valence-$\pi^*$, w/o CAP) & & & &\\ 
\hline
\end{tabular}
	
\end{table}

Overall, proposed form of the analytic gradients allows for geometry optimization on complex potential energy surfaces with a variety of electronic structure methods. Here we demonstrate the use of CASSCF and MR-CIS approaches. The reported optimized geometries of the metastable anions point to differences in the approximate localized treatment of the resonance state and CAP-based approaches, pointing to the necessity of appropriate treatment of interaction with the continuum.

\subsection{Conclusions}
The described formulation of the analytic nuclear gradients allows one to extend the methods and software for calculating gradients and non-adiabatic couplings for bound electronic states to resonances. The computed analytic gradients are in excellent agreement with their numerical counterparts, despite some terms have been neglected. In this work, we have presented implementation of analytic gradients for two projected CAP based electronic structure methods: CASSCF and MR-CIS. While CASSCF, as can be expected, yields poor energetic parameters owing to the lack of account for dynamic electron correlation, optimized geometries capture the qualitative changes observed with a more reliable MR-CIS method. The described scheme also produces non-adiabatic couplings as a by-product, \textit{i.e.} as off-diagonal matrix elements of the gradient matrix in the diagonal basis. The described developments pave the way for on-the-fly Born-Oppenheimer and non-adiabatic dynamics involving metastable states. 

\section{Supporting Information}
The Supporting Information contains information on the details of the numerical analysis of the presented pCAP gradient and on the comparison with other methods, presents energetic parameters of all discussed shape resonances, and lists diffuse basis sets used for calculation. The OpenCAP software with CAP gradient implementation is available in \url{https://github.com/SoubhikM/opencap/} with geometry optimization interfaces with COLUMBUS and OpenMolcas in \url{https://github.com/SoubhikM/OPENCAP-MD/}.

\section{Acknowledgments}
The authors thank Dr. Mushir Thodika and Professor Spiridoula Matsika for providing their MR-CIS potential energy curve data for $\mathrm{N_2}$. Simulations were performed on Boston University’s Shared Computing Cluster (SCC).

\bibliography{gradient_pcap}
\newpage
\input{SI_gradient_pcap}

\end{document}

%% file: SI_gradient_pcap.tex
\overfullrule=5pt 

\renewcommand{\thesection}{S\arabic{section}.}
\renewcommand{\thesubsection}{\thesection\arabic{subsection}.}
\renewcommand{\thesubsubsection}{S\thesubsection\arabic{subsubsection}.}

\renewcommand{\thefigure}{S\arabic{figure}}
\renewcommand{\thetable}{S\arabic{table}}
\renewcommand{\theequation}{S\arabic{equation}}

\renewcommand{\thepage}{S\arabic{page}}
\setcounter{section}{0}
\setcounter{figure}{0}
\setcounter{table}{0}
\setcounter{equation}{0}

\begin{center}
	{\it \Large Supporting Information\\
		Complex potential energy surfaces: gradients with projected CAP technique}\\

	Soubhik Mondal and Ksenia B. Bravaya \\
	Department of Chemistry, Boston University, Boston, Massachusetts, \nolinebreak{02215, USA}
\end{center}

\section{Comparison of CAP contributions to the gradient: CAP-EOM-EA-CCSD vs pCAP-EOM-EA-CCSD}
\begin{table}[!h]
	\centering
		\renewcommand{\arraystretch}{1.2} 
		\begin{tabular}{cccc|ccc}
            \toprule
			& \multicolumn{3}{c}{\textbf{CAP-EOM-EA-CCSD}} & \multicolumn{3}{c}{\textbf{pCAP-EOM-EA-CCSD}} \\
			\cmidrule(lr){2-4} \cmidrule(lr){5-7}
			& \multicolumn{3}{c}{$\mathbf{E_R^{\text{un-corr}}}$, $\boldsymbol{\Gamma}$ : (2.605, 0.542) eV} & \multicolumn{3}{c}{$\mathbf{E_R^{\text{un-corr}}}$, $\boldsymbol{\Gamma}$ : (2.605, 0.546) eV} \\
			\cmidrule(lr){2-4} \cmidrule(lr){5-7}
			\textbf{Atom} & \textbf{X}  & \textbf{Y} & \textbf{Z} & \textbf{X} & \textbf{Y} & \textbf{Z} \\
			\midrule
			N1 & 0.00000000 & 0.00000000 & ~0.0567769 & 0.00000000 & -0.00000000 & ~ 0.05810603 \\
			N2 & 0.00000000 & 0.00000000 & -0.0567769 & 0.00000000 & -0.00000000 & -0.05810603 \\
            \bottomrule
		\end{tabular}
	\caption{Comparison of CAP gradients $\mathrm{\left (\frac{hartree}{bohr} \right)}$ between CAP-EOM-EA-CCSD and pCAP-EOM-EA-CCSD. CAP strength $\eta_{opt}$=0.00137 and onset $r_{\alpha\{x, y, z\}}$=\{2.76, 2.76, 4.88\} bohr are the parameters chosen to produce the data at R(N-N) of 1.09 \AA.}
	\label{tab:merged_capgrad}
\end{table}

In Table \ref{tab:merged_capgrad}, we report CAP contributions to the gradient for a single point calculation of $\mathrm{N_2^-} (^2\pi_g)$ resonance in CAP-EOM-EA-CCSD\cite{benda2017capeom}\nocite{benda2017capeom} and pCAP-EOM-EA-CCSD\cite{gayvert2022projectedeom}\nocite{gayvert2022projectedeom} levels respectively.

The similarities in magnitudes and signs of the CAP gradients from two different methods underlines that firstly, complex 1-RDMs from our method \eqref{dm1ao_rotation} is similar to CAP-EOM-EA-CCSD where CAP is introduced in HF level and secondly, the evaluation of CAP derivative terms implementation on the grid is correct. We don't expect the complex  densities and CAP terms to be exactly same even though the complex energies are same because they are entirely different methods.

In `diag' basis, we evaluate the 1-particle reduce density matrices (1-RDM) as following,
\begin{equation}
	\begin{aligned}
		\mathbf{D^{\text{diag}}_{IJ}} &= \sum_{K=1}^\mathrm{N_{states}} \sum_{L=1}^\mathrm{N_{states}} \mathbf{C_{IK}^T} \, \mathbf{D^{\text{ZO}}_{KL}} \,\mathbf{C_{LJ}} \\
	\end{aligned}
	\label{dm1ao_rotation}
\end{equation}
,where $\mathbf{D^{\text{diag}}_{IJ}}$ and ,where $\mathbf{D^{\text{ZO}}_{IJ}}$ are the 1-TDM (transition density matrices) between state I and J in `diag' and ZO basis respectively. If I and J are the same, they correspond to state density matrices.

\section{Rotation vector contribution to the gradient}
We utilize the orthogonality relation of the rotation matrix $\mathbf{C}$: $\mathbf{CC^T=1}$ and the eigenvalue equation: $\mathbf{H^{CAP}C={E^\#}C}$, $\mathbf{C^TH^{CAP}={E^\#}C^T}$, where $\mathbf{E^\#}$ are the complex eigen-values. It follows that,
\begin{equation}
	\begin{aligned}
		\nabla_R \mathbf{C^T H^{CAP} C }+  \mathbf{C^T H^{CAP}} \nabla_R \mathbf{C}
		&= \mathbf{E^\#} \left[  \nabla_R \mathbf{C^T C} \right] + \mathbf{E^\#} \left[ \mathbf{C^T} \nabla_R \mathbf{C}\right]\\
		&= \mathbf{E^\#}  \nabla_R \left[\mathbf{ C^T C} \right] = 0\
	\end{aligned}
\end{equation}

\section{Tracking resonance state: using attachment and detachment densities}
The tracking algorithm we deploy when full wave-function overlap is not available, is through comparison of attachment and detachment densities\cite{head1995analysisADden, closser2014simulationstrack}\nocite{head1995analysisADden, closser2014simulationstrack} in subsequent steps, an approach which is also adopted in the Q-CHEM program\cite{qchem}\nocite{qchem}. Equation-wise, the overlap metric $\mathrm{S^{AD-DM, diag}_{IJ}}$ between state I and J from consecutive steps is evaluated as following\cite{closser2014simulationstrack}\nocite{closser2014simulationstrack},
\begin{equation}
		\begin{aligned}
	S^\text{AD-DM, diag}_{IJ} &= 1 - \frac{1}{2} \left|(A{'}_{I}^\text{diag}-A_{J}^\text{diag})-(D{'}^\text{diag}_{I}-D^\text{diag}_{J}) \right|
	\label{ovlp_dens_diag}
	\end{aligned}		
\end{equation}
where, $\mathrm{A^{diag}}$ and $\mathrm{D{'}^{diag}}$ are attachment and detachment densities in `diag' basis for old and new iterations respectively.  In equation~\eqref{ovlp_dens_diag}, the norm of the matrix M is evaluated as $\mathrm{\sqrt{\lambda_{max}}}$ with $\mathrm{\lambda}$ being eigenvalues of $\mathrm{MM^T}$. Essentially, equation~\eqref{ovlp_dens_diag} provides an estimate of how the density of a state changes between two iterations, if the densities are close, $\mathrm{S^{AD-DM, diag}_{IJ} }$ is $\sim$ 1.  

The attachment and detachment densities for state $I$ are calculated using prescriptions from reference \citenum{head1995analysisADden} and the one particle difference density matrix ($\Delta$)\cite{head1995analysisADden}\nocite{head1995analysisADden} is built with the reference state chosen as the neutral state evaluated in the same orthogonal basis to that of the correlated anionic states. The density contraction to `diag' from ZO basis is done using equation~\eqref{dm1ao_rotation}.

\newpage
\section{pCAP parameters for initial calculations at equilibrium geometries of the neutral \& energetics}
\begin{table}[!h]
	\label{tab:N2_eta_params}
	\small
	\centering
	\setlength{\tabcolsep}{9pt} 
	\renewcommand{\arraystretch}{1.4} 
	\begin{tabular}{lcccc}
		\toprule
		\textbf{Methods/basis} &  $\mathbf{r^0_x}$& $\mathbf{r^0_y}$  & $\mathbf{r^0_z}$ &  $\mathbf{\eta_{opt}}$  \\ 
		\midrule
		SA-10-CASSCF(5e, 13o) &2.77 & 2.77 & 4.88 &0.00790\\
		cc-pVTZ+[2s5p2d] & & & &\\ \hline
		MR-CIS/SA-6-CASSCF(7e, 10o) &2.77 & 2.77& 4.88&0.00854\\ 
		cc-pVTZ+[2s5p2d]& & & & \\ \hline
		MR-CIS/SA-6-CASSCF(5e, 9o) &2.77 & .2.77 &  4.88 &0.00724 \\ 
		cc-pVTZ+[2s5p2d]& & & & \\ \hline
		MR-CIS/SA-10-CASSCF(5e, 8o) &2.76 & 2.76 &  4.88 &0.00826 \\ 
		aug-cc-pVTZ+X[3s3p3d] & & & &\\ 
		\bottomrule
	\end{tabular}
	\caption{Parameters of $\mathrm{N_2^-}$: CAP onset $\mathrm{r^0_{\alpha=\left\{x,y,z\right\}}}$ and $\mathrm{\eta_{opt}}$ reported in atomic units for different method/basis sets.}		
\end{table}

\begin{table}[!h]
	\centering
	\renewcommand{\arraystretch}{1.2} 
		\begin{tabular}{lcccccc}
            \toprule
			\multirow{3}{*}{\textbf{Methods/basis}} & \multicolumn{4}{c}{ \textbf{$\mathbf{H_2CO (R_e)}$}} & \multicolumn{2}{c}{ \textbf{$\mathbf{H_2CO^- (R_e^{res}})$}} \\ 
			\cmidrule(lr){2-5} \cmidrule(lr){6-7}
			& $\mathbf{r^0_{\alpha=\left\{x,y,z\right\}}}$ & $\boldsymbol{\Delta E_0}$ & $\boldsymbol{\Gamma_0}$ &  $\mathbf{\eta_{opt}}$ & $\boldsymbol{\Delta E_{res}}$ & $\boldsymbol{\Gamma _{res}}$ \\ 
			\midrule
			SA-15-CASSCF(3e, 16o)/ & 5.98 & 2.426 & 0.338 & 0.01656  & 1.281 & 0.232 \\
			cc-pVTZ+X[4p] & 3.88& & & & & \\
			& 2.93& & & & & \\ \hline
			SA-19-CASSCF(3e, 20o)/ & 5.99 & 2.633 & 0.416 & 0.02888 & 1.086 & 0.687 \\
			cc-pVTZ+X[2s5p2d] & 3.89& & & & & \\ 
			& 2.94& & & & & \\ \hline
			MRCIS/SA-7-CASSCF(3e, 8o)/ & 6.05 &0.931 & 0.211 & 0.00196 & 0.168 & 0.101 \\
			cc-pVDZ+[3p]  (C$_s$ symmetry) & 3.94& & & & & \\ 
			& 2.97& & & & & \\
            \bottomrule
		\end{tabular}
	\caption{Parameters of CH$_2$O$^-$ : CAP onset $\mathrm{r^0_{\alpha=\left\{x,y,z\right\}}}$ and $\mathrm{\eta_{opt}}$ for different method/basis sets. $\mathrm{\Delta E_0}$ and $\mathrm{\Gamma_0}$ are the vertical energy gaps and decay widths  respectively calculated at equilibrium geometry of the neutral ($\mathrm{R_e}$). $\mathrm{\Delta E_{res}}$ and $\mathrm{\Gamma_{res}}$ are the vertical energy gaps and decay widths  respectively calculated at equilibrium geometry of the anion ($\mathrm{R_e^{res}}$). Energies are reported in eV and onset values in bohr.}
	\label{tab:H2CO_eta_params}
\end{table}

\begin{table}[!h]
	\centering
	\renewcommand{\arraystretch}{1.2} 
		\begin{tabular}{lcccccc}
            \toprule
			\multirow{3}{*}{\textbf{Methods/basis}} & \multicolumn{4}{c}{ \textbf{$\mathbf{C_2H_4 (R_e)}$}} & \multicolumn{2}{c}{ \textbf{$\mathbf{C_2H_4^- (R_e^{res}})$}} \\ 
			\cmidrule(lr){2-5} \cmidrule(lr){6-7}
			& $\mathbf{r^0_{\alpha=\left\{x,y,z\right\}}}$ & $\boldsymbol{\Delta E_0}$ & $\boldsymbol{\Gamma_0}$ &  $\mathbf{\eta_{opt}}$ & $\boldsymbol{\Delta E_{res}}$ & $\boldsymbol{\Gamma _{res}}$\\ 
			\midrule
			SA-15-CASSCF(3e, 20o)/ & 7.62 & 3.119 & 0.634 & 0.01772  & 2.350 & 0.364 \\
			cc-pVTZ+X[3p] & 4.20& & & & & \\
			& 6.02& & & & & \\ \hline
			MR-CIS/SA-6-CASSCF(3e, 7o)/ & 7.14 &2.617 & 0.592 & 0.00465 & 2.079& 0.348 \\
			cc-pVDZ+[3p] ($\mathrm{C_{2h}}$ symmetry) & 3.50& & & & & \\ 
			& 4.69& & & & & \\
            \bottomrule
		\end{tabular}
	\caption{Parameters of $\mathrm{C_2H_4^-}$ : CAP onset $\mathrm{r^0_{\alpha=\left\{x,y,z\right\}}}$ and $\mathrm{\eta_{opt}}$ for different method/basis sets. $\mathrm{\Delta E_0}$ and $\mathrm{\Gamma_0}$ are the vertical energy gaps and decay widths  respectively calculated at equilibrium geometry of the neutral ($\mathrm{R_e}$). $\mathrm{\Delta E_{res}}$ and $\mathrm{\Gamma_{res}}$ are the vertical energy gaps and decay widths  respectively calculated at equilibrium geometry of the anion ($\mathrm{R_e^{res}}$). Energies are reported in eV and onset values in bohr.}
	\label{tab:C2H4_eta_params}
\end{table}

\begin{table}[!h]
	\centering
	\renewcommand{\arraystretch}{1.2} 
		\begin{tabular}{lcccccc}
        \toprule
			\multirow{3}{*}{\textbf{Methods/basis}} & \multicolumn{4}{c}{ \textbf{$\mathbf{H_2CO_2 (R_e)}$}} & \multicolumn{2}{c}{ \textbf{$\mathbf{H_2CO_2^- (R_e^{res}})$}} \\ 
			\cmidrule(lr){2-5} \cmidrule(lr){6-7}
			& $\mathbf{r^0_{\alpha=\left\{x,y,z\right\}}}$ & $\boldsymbol{\Delta E_0}$ & $\boldsymbol{\Gamma_0}$ &  $\mathbf{\eta_{opt}}$ & $\boldsymbol{\Delta E_{res}}$ & $\boldsymbol{\Gamma _{res}}$ \\ 
			\midrule
			SA-16-CASSCF(3e, 17o)/ & 9.45 & 3.889 & 0.199 & 0.00743 & 1.591 & 0.258 \\
			cc-pVTZ+X[2s5p2d] & 5.29& & & & & \\ 
			& 3.56& & & & & \\ \hline
			SA-16-CASSCF(3e, 17o)/ & 9.45 & 4.116 & 0.240 & 0.00864& 1.580 & 1.125\\
			cc-pVTZ+X[3p] & 5.28& & & & & \\ 
			& 3.52& & & & & \\ \hline
			SA-15-CASSCF(3e, 23o)/ &9.54 & 3.739 & 0.065 & 0.04005 & 1.636 & 0.103 \\
			cc-pVTZ+X[4p] & 6.36& & & & & \\
			& 6.33& & & & & \\  \hline
			SA-15-CASSCF(3e, 23o)/ & 9.54 & 4.160 & 0.133 &0.03878& 1.674&0.496 \\
			cc-pVTZ+X[5p] & 5.29& & & & & \\ 
			& 3.56& & & & & \\ 
            \bottomrule
		\end{tabular}
	\caption{Parameters of $\mathrm{H_2CO_2^-}$ : CAP onset $\mathrm{r^0_{\alpha=\left\{x,y,z\right\}}}$ and $\mathrm{\eta_{opt}}$ for different method/basis sets. $\mathrm{\Delta E_0}$ and $\mathrm{\Gamma_0}$ are the vertical energy gaps and decay widths  respectively calculated at equilibrium geometry of the neutral ($\mathrm{R_e}$). $\mathrm{\Delta E_{res}}$ and $\mathrm{\Gamma_{res}}$ are the vertical energy gaps and decay widths  respectively calculated at equilibrium geometry of the anion ($\mathrm{R_e^{res}}$). Energies are reported in eV and onset values in bohr.}
	
	\label{tab:H2CO2_eta_params}
\end{table}

\newpage

\section{Augmented diffuse basis}
The main parent basis used in our calculations is cc-pVTZ, unless mentioned this is the case for the following.
\subsection{$\mathrm{N_2}$}
2s,2d scaling exponent of $\zeta_{i+1}=\frac{1}{2}\zeta_{i}$. 5p scaling exponent of $\zeta_{i+1}=\frac{2}{3}\zeta_{i}$.
\begin{verbatim}
	N			0
	S   1   1.00
	0.0893500  1.0000000
	S   1   1.00
	0.0446750  1.0000000
	P   1   1.00
	0.1150     1.0000000
	P   1   1.00
	0.0766667  1.0000000
	P   1   1.00
	0.0511111  1.0000000
	P   1   1.00
	0.0340741  1.0000000
	P   1   1.00
	0.0227160  1.0000000
	D   1   1.00
	0.234500   1.000000
	D   1   1.00
	0.117250   1.000000
\end{verbatim}

3s,3p.3d scaling exponent $\zeta_{i+1}=\frac{1}{2}\zeta_{i}$ for $\mathrm{N_2}$ for aug-cc-pVTZ.
\begin{verbatim}
	X		0
	S   1   1.00
	0.0288    1.00000000
	S   1   1.00
	0.0144    1.00000000
	S   1   1.00
	0.0072    1.00000000
	P   1   1.00
	0.02455    1.0000000	
	P   1   1.00
	0.012275   1.00000000
	P   1   1.00
	0.0061375  1.00000000 
	D   1   1.00
	0.0755        1.000000 
	D   1   1.00
	0.03775       1.00000
	D   1   1.00
	0.018875      1.00000
\end{verbatim}

\subsection{$\mathrm{H_2CO}$}
3p scaling exponent $\zeta_{i+1}=\frac{1}{2}\zeta_{i}$ with cc-pVDZ.
\begin{verbatim}
	C		0
	P    1   1.00
	0.13765     1.0
	P    1   1.00
	0.068825    1.0
	P    1   1.00
	0.0344125   1.0
\end{verbatim}

4p scaling exponent $\zeta_{i+1}=\frac{2}{3}\zeta_{i}$
\begin{verbatim}
	X     0
	P    1   1.00
	0.0806           1.000000D+00
	P    1   1.00
	0.05373           1.000000D+00
	P    1   1.00
	0.03582           1.000000D+00
	P    1   1.00
	0.02388	1.00
\end{verbatim}

2s,2d scaling exponent of $\zeta_{i+1}=\frac{1}{2}\zeta_{i}$. 5p scaling exponent of $\zeta_{i+1}=\frac{2}{3}\zeta_{i}$.
\begin{verbatim}
	X     0
	S    1   1.00
	0.05135	1.0         
	S    1   1.00
	0.025675              1.0000000
	P    1   1.00
	0.0806           1.000000D+00
	P    1   1.00
	0.05373           1.000000D+00
	P    1   1.00
	0.03582           1.000000D+00
	P    1   1.00
	0.02388	1.00
	P    1   1.00
	0.01592	1.00
	D    1   1.00
	0.159           1.0000000
	D    1   1.00
	0.0795           1.0000000
\end{verbatim}
\subsection{$\mathrm{C_2H_4}$}

3p scaling exponent $\zeta_{i+1}=\frac{1}{2}\zeta_{i}$ with cc-pVDZ.
\begin{verbatim}
	P    1   1.00
	0.13765     1.0
	P    1   1.00
	0.068825    1.0
	P    1   1.00
	0.0344125   1.0
\end{verbatim}

3p scaling exponent $\zeta_{i+1}=\frac{2}{3}\zeta_{i}$.
\begin{verbatim}
	C	0
	P   1   1.00
	0.0806	1.
	P   1   1.00
	0.05373333333333333	1.
	P   1   1.00
	0.03582222222222222	1.
\end{verbatim}

\subsection{$\mathrm{H_2CO_2}$}

4p/5p: scaling exponent of $\zeta_{i+1}=\frac{1}{2}\zeta_{i}$.
\begin{verbatim}
X     0
P    1   1.00
0.06045              1.000000D+00
P    1   1.00
0.030225            1.000000D+00
P    1   1.00
0.0151125           1.000000D+00
P    1   1.00
0.00755625	     1.000000D+00
P       1       1.00
0.003778125     1.000000D+00
\end{verbatim}

3p, scaling exponent $\zeta_{i+1}=\frac{2}{3}\zeta_{i}$.
\begin{verbatim}
X     0
P    1   1.00
0.0806             1.000000D+00
P    1   1.00
0.05373           1.000000D+00
P    1   1.00
0.03582           1.000000D+00
\end{verbatim}

2s,2d scaling exponent is $\zeta_{i+1}=\frac{1}{2}\zeta_{i}$). 5p scaling exponent is $\zeta_{i+1}=\frac{2}{3}\zeta_{i}$..
\begin{verbatim}
X     0
S    1   1.00
0.05135	1.0         
S    1   1.00
0.025675              1.0000000
P    1   1.00
0.0806           1.000000D+00
P    1   1.00
0.05373           1.000000D+00
P    1   1.00
0.03582           1.000000D+00
P    1   1.00
0.02388	1.0
P    1   1.00
0.01592	1.0
D    1   1.00
0.159           1.0000000
D    1   1.00
0.0795           1.0000000
\end{verbatim}

\section{Initial geometries}
\subsection{Initial geometry (in \AA) of $\mathrm{N_2}$ with cc-pVTZ basis}with $\mathrm{D_{2h}}$ symmetry.
\begin{verbatim}
 N     0.00000000    0.00000000    0.54743386 
 N     0.00000000    0.00000000   -0.54743386 
\end{verbatim}
\subsection{Initial geometry (in \AA) of $\mathrm{H_2CO}$ with cc-pVTZ basis}with $\mathrm{C_{1}}$ symmetry.
\begin{verbatim}
C   0.539081     -0.009227   0.000000
O  -0.636180     -0.086613   0.000000
H   1.181355     -0.893267   0.000000
H   1.056513      0.968385   0.000000
\end{verbatim}
\subsection{Initial geometry (in \AA) of $\mathrm{H_2CO}$ with cc-pVDZ basis}with $\mathrm{C_{s}}$ symmetry.
\begin{verbatim}
C   0.535543212514   -0.008891235494    0.00000000
O  -0.643690629882   -0.088924526013    0.00000000
H   1.185776060252   -0.897982505052    0.00000000
H   1.063138712502    0.975076216165    0.00000000
\end{verbatim}
\subsection{Initial geometry (in \AA) of $\mathrm{H_2CO_2^-}$ with cc-pVTZ basis}with $\mathrm{C_{1}}$ symmetry.
\begin{verbatim}
C   0.150321    -0.378974   0.000000
H   0.213778    -1.461673   0.000000
O   1.082107     0.341490   0.000000
O  -1.109733     0.010919   0.000000
H  -1.143257     0.957261   0.000000
\end{verbatim}
\subsection{Initial geometry (in \AA) of $\mathrm{H_2CO_2^-}$ with cc-pVDZ basis}with $\mathrm{C_{1}}$ symmetry.
\begin{verbatim}
C   0.151159     -0.375236   0.000000
H   0.213978     -1.480920   0.000000
O   1.079429      0.352747   0.000000
O  -1.113650      0.010337   0.000000
H  -1.137699      0.962095   0.000000
\end{verbatim}
\subsection{Initial geometry (in \AA) of $\mathrm{C_2H_4}$ with cc-pVTZ basis} with $\mathrm{C_{1}}$ symmetry.
\begin{verbatim}
C       -0.0179750928   -0.1136932028   -0.0115175428
H       -0.0496468803    0.0145973544    1.0544147734
H        0.9369204904    0.0146037045   -0.4862662659
C       -1.1084247201   -0.4170070183   -0.7097827982
H       -1.0767529326   -0.5452975755   -1.7757145852
H       -2.0633203033   -0.5453028673   -0.2350335460
\end{verbatim}
\subsection{Initial distorted geometry (in \AA) of $\mathrm{C_2H_4}$ with cc-pVTZ basis} with $\mathrm{C_{2h}}$ symmetry.
\begin{verbatim}
	C         0.670519        -0.000015          0.000000
	C        -0.670519         0.000015          0.000000
	H         1.238290         0.000271          0.922918
	H        -1.238290        -0.000271          0.922918
	H         1.238290         0.000271         -0.922918
	H        -1.238290        -0.000271         -0.922918
\end{verbatim}
\subsection{Initial distorted geometry (in \AA) of $\mathrm{C_2H_4}$ with cc-pVDZ basis} with $\mathrm{C_{2h}}$ symmetry.
\begin{verbatim}
	C         0.670518         0.000007          0.000000
	C        -0.670518        -0.000007          0.000000
	H         1.238283        -0.000164          0.922917
	H        -1.238283         0.000164          0.922917
	H         1.238283        -0.000164         -0.922917
	H        -1.238283         0.000164         -0.922917
\end{verbatim}
\newpage
\vfill

%% file: gradient_pcap.bbl
\providecommand{\latin}[1]{#1}
\makeatletter
\providecommand{\doi}
  {\begingroup\let\do\@makeother\dospecials
  \catcode`\{=1 \catcode`\}=2 \doi@aux}
\providecommand{\doi@aux}[1]{\endgroup\texttt{#1}}
\makeatother
\providecommand*\mcitethebibliography{\thebibliography}
\csname @ifundefined\endcsname{endmcitethebibliography}
  {\let\endmcitethebibliography\endthebibliography}{}
\begin{mcitethebibliography}{83}
\providecommand*\natexlab[1]{#1}
\providecommand*\mciteSetBstSublistMode[1]{}
\providecommand*\mciteSetBstMaxWidthForm[2]{}
\providecommand*\mciteBstWouldAddEndPuncttrue
  {\def\EndOfBibitem{\unskip.}}
\providecommand*\mciteBstWouldAddEndPunctfalse
  {\let\EndOfBibitem\relax}
\providecommand*\mciteSetBstMidEndSepPunct[3]{}
\providecommand*\mciteSetBstSublistLabelBeginEnd[3]{}
\providecommand*\EndOfBibitem{}
\mciteSetBstSublistMode{f}
\mciteSetBstMaxWidthForm{subitem}{(\alph{mcitesubitemcount})}
\mciteSetBstSublistLabelBeginEnd
  {\mcitemaxwidthsubitemform\space}
  {\relax}
  {\relax}

\bibitem[Boudaïffa \latin{et~al.}(2000)Boudaïffa, Cloutier, Hunting, Huels,
  and Sanche]{boudaiffa2000resonant}
Boudaïffa,~B.; Cloutier,~P.; Hunting,~D.; Huels,~M.~A.; Sanche,~L. Resonant
  formation of DNA strand breaks by low-energy (3 to 20 eV) electrons.
  \emph{Science} \textbf{2000}, \emph{287}, 1658--1660\relax
\mciteBstWouldAddEndPuncttrue
\mciteSetBstMidEndSepPunct{\mcitedefaultmidpunct}
{\mcitedefaultendpunct}{\mcitedefaultseppunct}\relax
\EndOfBibitem
\bibitem[Alizadeh \latin{et~al.}(2015)Alizadeh, Orlando, and
  Sanche]{alizadeh2015biomolecular}
Alizadeh,~E.; Orlando,~T.~M.; Sanche,~L. Biomolecular damage induced by
  ionizing radiation: the direct and indirect effects of low-energy electrons
  on DNA. \emph{Annual review of physical chemistry} \textbf{2015}, \emph{66},
  379--398\relax
\mciteBstWouldAddEndPuncttrue
\mciteSetBstMidEndSepPunct{\mcitedefaultmidpunct}
{\mcitedefaultendpunct}{\mcitedefaultseppunct}\relax
\EndOfBibitem
\bibitem[Tonzani and Greene(2006)Tonzani, and Greene]{tonzani2006low}
Tonzani,~S.; Greene,~C.~H. Low-energy electron scattering from DNA and RNA
  bases: Shape resonances and radiation damage. \emph{The Journal of chemical
  physics} \textbf{2006}, \emph{124}, 054312\relax
\mciteBstWouldAddEndPuncttrue
\mciteSetBstMidEndSepPunct{\mcitedefaultmidpunct}
{\mcitedefaultendpunct}{\mcitedefaultseppunct}\relax
\EndOfBibitem
\bibitem[Li \latin{et~al.}(2010)Li, Cloutier, Sanche, and Wagner]{li2010low}
Li,~Z.; Cloutier,~P.; Sanche,~L.; Wagner,~J.~R. Low-energy electron-induced DNA
  damage: Effect of base sequence in oligonucleotide trimers. \emph{Journal of
  the American Chemical Society} \textbf{2010}, \emph{132}, 5422--5427\relax
\mciteBstWouldAddEndPuncttrue
\mciteSetBstMidEndSepPunct{\mcitedefaultmidpunct}
{\mcitedefaultendpunct}{\mcitedefaultseppunct}\relax
\EndOfBibitem
\bibitem[Mason \latin{et~al.}(2014)Mason, Nair, Jheeta, and
  Szyma{\'n}ska]{mason2014electron}
Mason,~N.~J.; Nair,~B.; Jheeta,~S.; Szyma{\'n}ska,~E. Electron induced
  chemistry: a new frontier in astrochemistry. \emph{Faraday discussions}
  \textbf{2014}, \emph{168}, 235--247\relax
\mciteBstWouldAddEndPuncttrue
\mciteSetBstMidEndSepPunct{\mcitedefaultmidpunct}
{\mcitedefaultendpunct}{\mcitedefaultseppunct}\relax
\EndOfBibitem
\bibitem[Boyer \latin{et~al.}(2016)Boyer, Rivas, Tran, Verish, and
  Arumainayagam]{boyer2016role}
Boyer,~M.~C.; Rivas,~N.; Tran,~A.~A.; Verish,~C.~A.; Arumainayagam,~C.~R. The
  role of low-energy ($\le$ 20 eV) electrons in astrochemistry. \emph{Surface
  Science} \textbf{2016}, \emph{652}, 26--32\relax
\mciteBstWouldAddEndPuncttrue
\mciteSetBstMidEndSepPunct{\mcitedefaultmidpunct}
{\mcitedefaultendpunct}{\mcitedefaultseppunct}\relax
\EndOfBibitem
\bibitem[Petrie and Bohme(2007)Petrie, and Bohme]{petrie2007ions}
Petrie,~S.; Bohme,~D.~K. Ions in space. \emph{Mass spectrometry reviews}
  \textbf{2007}, \emph{26}, 258--280\relax
\mciteBstWouldAddEndPuncttrue
\mciteSetBstMidEndSepPunct{\mcitedefaultmidpunct}
{\mcitedefaultendpunct}{\mcitedefaultseppunct}\relax
\EndOfBibitem
\bibitem[Millar \latin{et~al.}(2017)Millar, Walsh, and
  Field]{millar2017negative}
Millar,~T.~J.; Walsh,~C.; Field,~T.~A. Negative ions in space. \emph{Chemical
  reviews} \textbf{2017}, \emph{117}, 1765--1795\relax
\mciteBstWouldAddEndPuncttrue
\mciteSetBstMidEndSepPunct{\mcitedefaultmidpunct}
{\mcitedefaultendpunct}{\mcitedefaultseppunct}\relax
\EndOfBibitem
\bibitem[Simons(2008)]{simons2008molecular}
Simons,~J. Molecular anions. \emph{The Journal of Physical Chemistry A}
  \textbf{2008}, \emph{112}, 6401--6511\relax
\mciteBstWouldAddEndPuncttrue
\mciteSetBstMidEndSepPunct{\mcitedefaultmidpunct}
{\mcitedefaultendpunct}{\mcitedefaultseppunct}\relax
\EndOfBibitem
\bibitem[Jordan and Burrow(1987)Jordan, and Burrow]{jordan1987temporary}
Jordan,~K.~D.; Burrow,~P.~D. Temporary anion states of polyatomic hydrocarbons.
  \emph{Chemical Reviews} \textbf{1987}, \emph{87}, 557--588\relax
\mciteBstWouldAddEndPuncttrue
\mciteSetBstMidEndSepPunct{\mcitedefaultmidpunct}
{\mcitedefaultendpunct}{\mcitedefaultseppunct}\relax
\EndOfBibitem
\bibitem[Taylor \latin{et~al.}(1966)Taylor, Nazaroff, and
  Golebiewski]{taylor1966qualitative}
Taylor,~H.~S.; Nazaroff,~G.~V.; Golebiewski,~A. Qualitative Aspects of
  Resonances in Electron—Atom and Electron—Molecule Scattering, Excitation,
  and Reactions. \emph{The Journal of Chemical Physics} \textbf{1966},
  \emph{45}, 2872--2888\relax
\mciteBstWouldAddEndPuncttrue
\mciteSetBstMidEndSepPunct{\mcitedefaultmidpunct}
{\mcitedefaultendpunct}{\mcitedefaultseppunct}\relax
\EndOfBibitem
\bibitem[Simons and Jordan(1987)Simons, and Jordan]{simons1987ab}
Simons,~J.; Jordan,~K.~D. Ab initio electronic structure of anions.
  \emph{Chemical Reviews} \textbf{1987}, \emph{87}, 535--555\relax
\mciteBstWouldAddEndPuncttrue
\mciteSetBstMidEndSepPunct{\mcitedefaultmidpunct}
{\mcitedefaultendpunct}{\mcitedefaultseppunct}\relax
\EndOfBibitem
\bibitem[Allan(1985)]{Allan_expt}
Allan,~M. Excitation of vibrational levels up to $\nu$=17 in N$_2$by electron
  impact in the 0-5 {eV} region. \emph{Journal of Physics B: Atomic and
  Molecular Physics} \textbf{1985}, \emph{18}, 4511--4517\relax
\mciteBstWouldAddEndPuncttrue
\mciteSetBstMidEndSepPunct{\mcitedefaultmidpunct}
{\mcitedefaultendpunct}{\mcitedefaultseppunct}\relax
\EndOfBibitem
\bibitem[Itikawa(2006)]{itikawa_expt}
Itikawa,~Y. Cross Sections for Electron Collisions with Nitrogen Molecules.
  \emph{Journal of Physical and Chemical Reference Data} \textbf{2006},
  \emph{35}, 31--53\relax
\mciteBstWouldAddEndPuncttrue
\mciteSetBstMidEndSepPunct{\mcitedefaultmidpunct}
{\mcitedefaultendpunct}{\mcitedefaultseppunct}\relax
\EndOfBibitem
\bibitem[Vicic \latin{et~al.}(Mar 1996)Vicic, Poparic, and Belic]{VicicMar1996}
Vicic,~M.; Poparic,~G.; Belic,~D.~S. Large vibrational excitation of N$_2$ by
  low-energy electrons. \emph{Journal of Physics B, Atomic, Molecular and
  Optical Physics} \textbf{Mar 1996}, \emph{29}, 1273--1281\relax
\mciteBstWouldAddEndPuncttrue
\mciteSetBstMidEndSepPunct{\mcitedefaultmidpunct}
{\mcitedefaultendpunct}{\mcitedefaultseppunct}\relax
\EndOfBibitem
\bibitem[Bald \latin{et~al.}(2008)Bald, Langer, Tegeder, and
  Ing{\'o}lfsson]{bald2008isolated}
Bald,~I.; Langer,~J.; Tegeder,~P.; Ing{\'o}lfsson,~O. From isolated molecules
  through clusters and condensates to the building blocks of life.
  \emph{International Journal of Mass Spectrometry} \textbf{2008}, \emph{277},
  4--25\relax
\mciteBstWouldAddEndPuncttrue
\mciteSetBstMidEndSepPunct{\mcitedefaultmidpunct}
{\mcitedefaultendpunct}{\mcitedefaultseppunct}\relax
\EndOfBibitem
\bibitem[Arumainayagam \latin{et~al.}(2010)Arumainayagam, Lee, Nelson, Haines,
  and Gunawardane]{arumainayagam2010low}
Arumainayagam,~C.~R.; Lee,~H.-L.; Nelson,~R.~B.; Haines,~D.~R.;
  Gunawardane,~R.~P. Low-energy electron-induced reactions in condensed matter.
  \emph{Surface Science Reports} \textbf{2010}, \emph{65}, 1--44\relax
\mciteBstWouldAddEndPuncttrue
\mciteSetBstMidEndSepPunct{\mcitedefaultmidpunct}
{\mcitedefaultendpunct}{\mcitedefaultseppunct}\relax
\EndOfBibitem
\bibitem[Bass and Sanche(2003)Bass, and Sanche]{bass2003dissociative}
Bass,~A.~D.; Sanche,~L. Dissociative electron attachment and charge transfer in
  condensed matter. \emph{Radiation Physics and Chemistry} \textbf{2003},
  \emph{68}, 3--13\relax
\mciteBstWouldAddEndPuncttrue
\mciteSetBstMidEndSepPunct{\mcitedefaultmidpunct}
{\mcitedefaultendpunct}{\mcitedefaultseppunct}\relax
\EndOfBibitem
\bibitem[Sch{\"u}rmann \latin{et~al.}(2017)Sch{\"u}rmann, Tsering, Tanzer,
  Denifl, Kumar, and Bald]{schurmann2017resonant}
Sch{\"u}rmann,~R.; Tsering,~T.; Tanzer,~K.; Denifl,~S.; Kumar,~S.; Bald,~I.
  Resonant Formation of Strand Breaks in Sensitized Oligonucleotides Induced by
  Low-Energy Electrons (0.5--9 eV). \emph{Angewandte Chemie International
  Edition} \textbf{2017}, \emph{56}, 10952--10955\relax
\mciteBstWouldAddEndPuncttrue
\mciteSetBstMidEndSepPunct{\mcitedefaultmidpunct}
{\mcitedefaultendpunct}{\mcitedefaultseppunct}\relax
\EndOfBibitem
\bibitem[Santra and Cederbaum(2002)Santra, and Cederbaum]{santra2002non}
Santra,~R.; Cederbaum,~L.~S. Non-Hermitian electronic theory and applications
  to clusters. \emph{Physics reports} \textbf{2002}, \emph{368}, 1--117\relax
\mciteBstWouldAddEndPuncttrue
\mciteSetBstMidEndSepPunct{\mcitedefaultmidpunct}
{\mcitedefaultendpunct}{\mcitedefaultseppunct}\relax
\EndOfBibitem
\bibitem[Santra \latin{et~al.}(2000)Santra, Zobeley, Cederbaum, and
  Moiseyev]{santra2000interatomic}
Santra,~R.; Zobeley,~J.; Cederbaum,~L.~S.; Moiseyev,~N. Interatomic Coulombic
  decay in van der Waals clusters and impact of nuclear motion. \emph{Physical
  review letters} \textbf{2000}, \emph{85}, 4490\relax
\mciteBstWouldAddEndPuncttrue
\mciteSetBstMidEndSepPunct{\mcitedefaultmidpunct}
{\mcitedefaultendpunct}{\mcitedefaultseppunct}\relax
\EndOfBibitem
\bibitem[Moiseyev \latin{et~al.}(2001)Moiseyev, Santra, Zobeley, and
  Cederbaum]{vib_fingerprint}
Moiseyev,~N.; Santra,~R.; Zobeley,~J.; Cederbaum,~L.~S. Fingerprints of the
  nodal structure of autoionizing vibrational wave functions in clusters:
  Interatomic Coulombic decay in Ne dimer. \emph{The Journal of Chemical
  Physics} \textbf{2001}, \emph{114}, 7351--7360\relax
\mciteBstWouldAddEndPuncttrue
\mciteSetBstMidEndSepPunct{\mcitedefaultmidpunct}
{\mcitedefaultendpunct}{\mcitedefaultseppunct}\relax
\EndOfBibitem
\bibitem[Moiseyev(1998)]{moiseyev1998quantum}
Moiseyev,~N. Quantum theory of resonances: calculating energies, widths and
  cross-sections by complex scaling. \emph{Physics reports} \textbf{1998},
  \emph{302}, 212--293\relax
\mciteBstWouldAddEndPuncttrue
\mciteSetBstMidEndSepPunct{\mcitedefaultmidpunct}
{\mcitedefaultendpunct}{\mcitedefaultseppunct}\relax
\EndOfBibitem
\bibitem[Domcke(1991)]{domcke1991theory}
Domcke,~W. Theory of resonance and threshold effects in electron-molecule
  collisions: The projection-operator approach. \emph{Physics reports}
  \textbf{1991}, \emph{208}, 97--188\relax
\mciteBstWouldAddEndPuncttrue
\mciteSetBstMidEndSepPunct{\mcitedefaultmidpunct}
{\mcitedefaultendpunct}{\mcitedefaultseppunct}\relax
\EndOfBibitem
\bibitem[Langhoff(1979)]{langhoff1979stieltjes}
Langhoff,~P. \emph{Electron-Molecule and Photon-Molecule Collisions}; Springer,
  1979; pp 183--224\relax
\mciteBstWouldAddEndPuncttrue
\mciteSetBstMidEndSepPunct{\mcitedefaultmidpunct}
{\mcitedefaultendpunct}{\mcitedefaultseppunct}\relax
\EndOfBibitem
\bibitem[Mandelshtam and Taylor(1995)Mandelshtam, and
  Taylor]{mandelshtam1995spectral}
Mandelshtam,~V.~A.; Taylor,~H.~S. Spectral projection approach to the quantum
  scattering calculations. \emph{The Journal of chemical physics}
  \textbf{1995}, \emph{102}, 7390--7399\relax
\mciteBstWouldAddEndPuncttrue
\mciteSetBstMidEndSepPunct{\mcitedefaultmidpunct}
{\mcitedefaultendpunct}{\mcitedefaultseppunct}\relax
\EndOfBibitem
\bibitem[Hazi and Taylor(1970)Hazi, and Taylor]{hazi1970stabilization}
Hazi,~A.~U.; Taylor,~H.~S. Stabilization method of calculating resonance
  energies: Model problem. \emph{Physical Review A} \textbf{1970}, \emph{1},
  1109\relax
\mciteBstWouldAddEndPuncttrue
\mciteSetBstMidEndSepPunct{\mcitedefaultmidpunct}
{\mcitedefaultendpunct}{\mcitedefaultseppunct}\relax
\EndOfBibitem
\bibitem[Jagau \latin{et~al.}(2017)Jagau, Bravaya, and
  Krylov]{jagau2017extending}
Jagau,~T.-C.; Bravaya,~K.~B.; Krylov,~A.~I. Extending quantum chemistry of
  bound states to electronic resonances. \emph{Annual review of physical
  chemistry} \textbf{2017}, \emph{68}, 525--553\relax
\mciteBstWouldAddEndPuncttrue
\mciteSetBstMidEndSepPunct{\mcitedefaultmidpunct}
{\mcitedefaultendpunct}{\mcitedefaultseppunct}\relax
\EndOfBibitem
\bibitem[Jagau and Krylov(2016)Jagau, and Krylov]{jagau2016characterizing}
Jagau,~T.-C.; Krylov,~A.~I. Characterizing metastable states beyond energies
  and lifetimes: Dyson orbitals and transition dipole moments. \emph{The
  Journal of chemical physics} \textbf{2016}, \emph{144}\relax
\mciteBstWouldAddEndPuncttrue
\mciteSetBstMidEndSepPunct{\mcitedefaultmidpunct}
{\mcitedefaultendpunct}{\mcitedefaultseppunct}\relax
\EndOfBibitem
\bibitem[Jagau(2022)]{jagau2022theory}
Jagau,~T.-C. Theory of electronic resonances: fundamental aspects and recent
  advances. \emph{Chemical Communications} \textbf{2022}, \emph{58},
  5205--5224\relax
\mciteBstWouldAddEndPuncttrue
\mciteSetBstMidEndSepPunct{\mcitedefaultmidpunct}
{\mcitedefaultendpunct}{\mcitedefaultseppunct}\relax
\EndOfBibitem
\bibitem[Moiseyev(2011)]{moiseyev2011non}
Moiseyev,~N. \emph{Non-Hermitian quantum mechanics}; Cambridge University
  Press, 2011\relax
\mciteBstWouldAddEndPuncttrue
\mciteSetBstMidEndSepPunct{\mcitedefaultmidpunct}
{\mcitedefaultendpunct}{\mcitedefaultseppunct}\relax
\EndOfBibitem
\bibitem[Gamow(1928)]{gamow1928quantentheorie}
Gamow,~G. Zur quantentheorie des atomkernes. \emph{Zeitschrift f{\"u}r Physik}
  \textbf{1928}, \emph{51}, 204--212\relax
\mciteBstWouldAddEndPuncttrue
\mciteSetBstMidEndSepPunct{\mcitedefaultmidpunct}
{\mcitedefaultendpunct}{\mcitedefaultseppunct}\relax
\EndOfBibitem
\bibitem[Siegert(1939)]{siegert1939derivation}
Siegert,~A.~J. On the derivation of the dispersion formula for nuclear
  reactions. \emph{Physical Review} \textbf{1939}, \emph{56}, 750\relax
\mciteBstWouldAddEndPuncttrue
\mciteSetBstMidEndSepPunct{\mcitedefaultmidpunct}
{\mcitedefaultendpunct}{\mcitedefaultseppunct}\relax
\EndOfBibitem
\bibitem[Gyamfi and Jagau(2022)Gyamfi, and Jagau]{gyamfi2022ab}
Gyamfi,~J.~A.; Jagau,~T.-C. Ab initio molecular dynamics of temporary anions
  using complex absorbing potentials. \emph{The Journal of Physical Chemistry
  Letters} \textbf{2022}, \emph{13}, 8477--8483\relax
\mciteBstWouldAddEndPuncttrue
\mciteSetBstMidEndSepPunct{\mcitedefaultmidpunct}
{\mcitedefaultendpunct}{\mcitedefaultseppunct}\relax
\EndOfBibitem
\bibitem[Kossoski and Barbatti(2020)Kossoski, and
  Barbatti]{kossoski2020nonadiabatic}
Kossoski,~F.; Barbatti,~M. Nonadiabatic dynamics in multidimensional complex
  potential energy surfaces. \emph{Chemical Science} \textbf{2020}, \emph{11},
  9827--9835\relax
\mciteBstWouldAddEndPuncttrue
\mciteSetBstMidEndSepPunct{\mcitedefaultmidpunct}
{\mcitedefaultendpunct}{\mcitedefaultseppunct}\relax
\EndOfBibitem
\bibitem[Moiseyev(2017)]{Moiseyev2017}
Moiseyev,~N. Forces on nuclei moving on autoionizing molecular potential energy
  surfaces. \emph{The Journal of Chemical Physics} \textbf{2017}, \emph{146},
  024101\relax
\mciteBstWouldAddEndPuncttrue
\mciteSetBstMidEndSepPunct{\mcitedefaultmidpunct}
{\mcitedefaultendpunct}{\mcitedefaultseppunct}\relax
\EndOfBibitem
\bibitem[Benda and Jagau(2017)Benda, and Jagau]{benda2017communication}
Benda,~Z.; Jagau,~T.-C. Communication: Analytic gradients for the complex
  absorbing potential equation-of-motion coupled-cluster method. \emph{The
  Journal of chemical physics} \textbf{2017}, \emph{146}\relax
\mciteBstWouldAddEndPuncttrue
\mciteSetBstMidEndSepPunct{\mcitedefaultmidpunct}
{\mcitedefaultendpunct}{\mcitedefaultseppunct}\relax
\EndOfBibitem
\bibitem[Ehara and Sommerfeld(2012)Ehara, and Sommerfeld]{ehara2012cap}
Ehara,~M.; Sommerfeld,~T. CAP/SAC-CI method for calculating resonance states of
  metastable anions. \emph{Chemical Physics Letters} \textbf{2012}, \emph{537},
  107--112\relax
\mciteBstWouldAddEndPuncttrue
\mciteSetBstMidEndSepPunct{\mcitedefaultmidpunct}
{\mcitedefaultendpunct}{\mcitedefaultseppunct}\relax
\EndOfBibitem
\bibitem[Gayvert and Bravaya(2022)Gayvert, and Bravaya]{gayvert2022projected}
Gayvert,~J.~R.; Bravaya,~K.~B. Projected CAP-EOM-CCSD method for electronic
  resonances. \emph{The Journal of Chemical Physics} \textbf{2022},
  \emph{156}\relax
\mciteBstWouldAddEndPuncttrue
\mciteSetBstMidEndSepPunct{\mcitedefaultmidpunct}
{\mcitedefaultendpunct}{\mcitedefaultseppunct}\relax
\EndOfBibitem
\bibitem[Kanazawa \latin{et~al.}(2016)Kanazawa, Ehara, and
  Sommerfeld]{kanazawa2016low}
Kanazawa,~Y.; Ehara,~M.; Sommerfeld,~T. Low-lying $\pi$* resonances of standard
  and rare DNA and RNA bases studied by the projected CAP/SAC--CI method.
  \emph{The Journal of Physical Chemistry A} \textbf{2016}, \emph{120},
  1545--1553\relax
\mciteBstWouldAddEndPuncttrue
\mciteSetBstMidEndSepPunct{\mcitedefaultmidpunct}
{\mcitedefaultendpunct}{\mcitedefaultseppunct}\relax
\EndOfBibitem
\bibitem[Thodika and Matsika(2022)Thodika, and Matsika]{thodika2022projected}
Thodika,~M.; Matsika,~S. \emph{J. Chem. Theory Comput.} \textbf{2022},
  \emph{18}, 3377--3390\relax
\mciteBstWouldAddEndPuncttrue
\mciteSetBstMidEndSepPunct{\mcitedefaultmidpunct}
{\mcitedefaultendpunct}{\mcitedefaultseppunct}\relax
\EndOfBibitem
\bibitem[Jagau and Krylov(2014)Jagau, and Krylov]{jagau2014complex}
Jagau,~T.-C.; Krylov,~A.~I. Complex absorbing potential equation-of-motion
  coupled-cluster method yields smooth and internally consistent potential
  energy surfaces and lifetimes for molecular resonances. \emph{The Journal of
  Physical Chemistry Letters} \textbf{2014}, \emph{5}, 3078--3085\relax
\mciteBstWouldAddEndPuncttrue
\mciteSetBstMidEndSepPunct{\mcitedefaultmidpunct}
{\mcitedefaultendpunct}{\mcitedefaultseppunct}\relax
\EndOfBibitem
\bibitem[Mondal and Bravaya(2024)Mondal, and Bravaya]{mondal2024complex}
Mondal,~S.; Bravaya,~K.~B. Complex potential energy surfaces with projected CAP
  technique: Vibrational excitation of N2. \emph{The Journal of Chemical
  Physics} \textbf{2024}, \emph{161}\relax
\mciteBstWouldAddEndPuncttrue
\mciteSetBstMidEndSepPunct{\mcitedefaultmidpunct}
{\mcitedefaultendpunct}{\mcitedefaultseppunct}\relax
\EndOfBibitem
\bibitem[MacLeod and Shiozaki(2015)MacLeod, and
  Shiozaki]{macleod2015communication}
MacLeod,~M.~K.; Shiozaki,~T. Communication: Automatic code generation enables
  nuclear gradient computations for fully internally contracted multireference
  theory. \emph{The Journal of Chemical Physics} \textbf{2015},
  \emph{142}\relax
\mciteBstWouldAddEndPuncttrue
\mciteSetBstMidEndSepPunct{\mcitedefaultmidpunct}
{\mcitedefaultendpunct}{\mcitedefaultseppunct}\relax
\EndOfBibitem
\bibitem[Vlaisavljevich and Shiozaki(2016)Vlaisavljevich, and
  Shiozaki]{vlaisavljevich2016nuclear}
Vlaisavljevich,~B.; Shiozaki,~T. Nuclear energy gradients for internally
  contracted complete active space second-order perturbation theory: Multistate
  extensions. \emph{Journal of chemical theory and computation} \textbf{2016},
  \emph{12}, 3781--3787\relax
\mciteBstWouldAddEndPuncttrue
\mciteSetBstMidEndSepPunct{\mcitedefaultmidpunct}
{\mcitedefaultendpunct}{\mcitedefaultseppunct}\relax
\EndOfBibitem
\bibitem[Park and Shiozaki(2017)Park, and Shiozaki]{park2017analytical}
Park,~J.~W.; Shiozaki,~T. Analytical derivative coupling for multistate CASPT2
  theory. \emph{Journal of chemical theory and computation} \textbf{2017},
  \emph{13}, 2561--2570\relax
\mciteBstWouldAddEndPuncttrue
\mciteSetBstMidEndSepPunct{\mcitedefaultmidpunct}
{\mcitedefaultendpunct}{\mcitedefaultseppunct}\relax
\EndOfBibitem
\bibitem[Park and Shiozaki(2017)Park, and Shiozaki]{park2017fly}
Park,~J.~W.; Shiozaki,~T. On-the-fly CASPT2 surface-hopping dynamics.
  \emph{Journal of chemical theory and computation} \textbf{2017}, \emph{13},
  3676--3683\relax
\mciteBstWouldAddEndPuncttrue
\mciteSetBstMidEndSepPunct{\mcitedefaultmidpunct}
{\mcitedefaultendpunct}{\mcitedefaultseppunct}\relax
\EndOfBibitem
\bibitem[Riss and Meyer(1993)Riss, and Meyer]{riss1993calculation}
Riss,~U.; Meyer,~H.-D. Calculation of resonance energies and widths using the
  complex absorbing potential method. \emph{Journal of Physics B: Atomic,
  Molecular and Optical Physics} \textbf{1993}, \emph{26}, 4503\relax
\mciteBstWouldAddEndPuncttrue
\mciteSetBstMidEndSepPunct{\mcitedefaultmidpunct}
{\mcitedefaultendpunct}{\mcitedefaultseppunct}\relax
\EndOfBibitem
\bibitem[Santra \latin{et~al.}(1999)Santra, Cederbaum, and
  Meyer]{santra1999electronic}
Santra,~R.; Cederbaum,~L.~S.; Meyer,~H.-D. Electronic decay of molecular
  clusters: non-stationary states computed by standard quantum chemistry
  methods. \emph{Chemical physics letters} \textbf{1999}, \emph{303},
  413--419\relax
\mciteBstWouldAddEndPuncttrue
\mciteSetBstMidEndSepPunct{\mcitedefaultmidpunct}
{\mcitedefaultendpunct}{\mcitedefaultseppunct}\relax
\EndOfBibitem
\bibitem[Sommerfeld and Ehara(2015)Sommerfeld, and
  Ehara]{sommerfeld2015complex}
Sommerfeld,~T.; Ehara,~M. Complex absorbing potentials with Voronoi isosurfaces
  wrapping perfectly around molecules. \emph{Journal of Chemical Theory and
  Computation} \textbf{2015}, \emph{11}, 4627--4633\relax
\mciteBstWouldAddEndPuncttrue
\mciteSetBstMidEndSepPunct{\mcitedefaultmidpunct}
{\mcitedefaultendpunct}{\mcitedefaultseppunct}\relax
\EndOfBibitem
\bibitem[Ehara \latin{et~al.}(2016)Ehara, Fukuda, and
  Sommerfeld]{ehara2016projected}
Ehara,~M.; Fukuda,~R.; Sommerfeld,~T. \emph{J. Comput. Chem.} \textbf{2016},
  \emph{37}, 242--249\relax
\mciteBstWouldAddEndPuncttrue
\mciteSetBstMidEndSepPunct{\mcitedefaultmidpunct}
{\mcitedefaultendpunct}{\mcitedefaultseppunct}\relax
\EndOfBibitem
\bibitem[Kunitsa \latin{et~al.}(2017)Kunitsa, Granovsky, and
  Bravaya]{kunitsa2017cap}
Kunitsa,~A.~A.; Granovsky,~A.~A.; Bravaya,~K.~B. \emph{J. Chem. Phys.}
  \textbf{2017}, \emph{146}\relax
\mciteBstWouldAddEndPuncttrue
\mciteSetBstMidEndSepPunct{\mcitedefaultmidpunct}
{\mcitedefaultendpunct}{\mcitedefaultseppunct}\relax
\EndOfBibitem
\bibitem[Phung \latin{et~al.}(2020)Phung, Komori, Yanai, Sommerfeld, and
  Ehara]{phung2020combination}
Phung,~Q.~M.; Komori,~Y.; Yanai,~T.; Sommerfeld,~T.; Ehara,~M. Combination of a
  Voronoi-type complex absorbing potential with the XMS-CASPT2 method and pilot
  applications. \emph{Journal of Chemical Theory and Computation}
  \textbf{2020}, \emph{16}, 2606--2616\relax
\mciteBstWouldAddEndPuncttrue
\mciteSetBstMidEndSepPunct{\mcitedefaultmidpunct}
{\mcitedefaultendpunct}{\mcitedefaultseppunct}\relax
\EndOfBibitem
\bibitem[Bravaya \latin{et~al.}(2013)Bravaya, Zuev, Epifanovsky, and
  Krylov]{bravaya2013complex}
Bravaya,~K.~B.; Zuev,~D.; Epifanovsky,~E.; Krylov,~A.~I. Complex-scaled
  equation-of-motion coupled-cluster method with single and double
  substitutions for autoionizing excited states: Theory, implementation, and
  examples. \emph{The Journal of chemical physics} \textbf{2013},
  \emph{138}\relax
\mciteBstWouldAddEndPuncttrue
\mciteSetBstMidEndSepPunct{\mcitedefaultmidpunct}
{\mcitedefaultendpunct}{\mcitedefaultseppunct}\relax
\EndOfBibitem
\bibitem[Zuev \latin{et~al.}(2014)Zuev, Jagau, Bravaya, Epifanovsky, Shao,
  Sundstrom, Head-Gordon, and Krylov]{zuev2014complex}
Zuev,~D.; Jagau,~T.-C.; Bravaya,~K.~B.; Epifanovsky,~E.; Shao,~Y.;
  Sundstrom,~E.; Head-Gordon,~M.; Krylov,~A.~I. Complex absorbing potentials
  within EOM-CC family of methods: Theory, implementation, and benchmarks.
  \emph{The Journal of chemical physics} \textbf{2014}, \emph{141}\relax
\mciteBstWouldAddEndPuncttrue
\mciteSetBstMidEndSepPunct{\mcitedefaultmidpunct}
{\mcitedefaultendpunct}{\mcitedefaultseppunct}\relax
\EndOfBibitem
\bibitem[White \latin{et~al.}(2017)White, Epifanovsky, McCurdy, and
  Head-Gordon]{white2017second}
White,~A.~F.; Epifanovsky,~E.; McCurdy,~C.~W.; Head-Gordon,~M. Second order
  M{\o}ller-Plesset and coupled cluster singles and doubles methods with
  complex basis functions for resonances in electron-molecule scattering.
  \emph{The Journal of Chemical Physics} \textbf{2017}, \emph{146}\relax
\mciteBstWouldAddEndPuncttrue
\mciteSetBstMidEndSepPunct{\mcitedefaultmidpunct}
{\mcitedefaultendpunct}{\mcitedefaultseppunct}\relax
\EndOfBibitem
\bibitem[Ghosh \latin{et~al.}(2012)Ghosh, Vaval, and Pal]{ghosh2012equation}
Ghosh,~A.; Vaval,~N.; Pal,~S. Equation-of-motion coupled-cluster method for the
  study of shape resonance. \emph{The Journal of Chemical Physics}
  \textbf{2012}, \emph{136}\relax
\mciteBstWouldAddEndPuncttrue
\mciteSetBstMidEndSepPunct{\mcitedefaultmidpunct}
{\mcitedefaultendpunct}{\mcitedefaultseppunct}\relax
\EndOfBibitem
\bibitem[Santra and Cederbaum(2002)Santra, and Cederbaum]{santra2002complex}
Santra,~R.; Cederbaum,~L.~S. Complex absorbing potentials in the framework of
  electron propagator theory. I. General formalism. \emph{The Journal of
  chemical physics} \textbf{2002}, \emph{117}, 5511--5521\relax
\mciteBstWouldAddEndPuncttrue
\mciteSetBstMidEndSepPunct{\mcitedefaultmidpunct}
{\mcitedefaultendpunct}{\mcitedefaultseppunct}\relax
\EndOfBibitem
\bibitem[Feuerbacher \latin{et~al.}(2003)Feuerbacher, Sommerfeld, Santra, and
  Cederbaum]{feuerbacher2003complex}
Feuerbacher,~S.; Sommerfeld,~T.; Santra,~R.; Cederbaum,~L.~S. Complex absorbing
  potentials in the framework of electron propagator theory. II. Application to
  temporary anions. \emph{The Journal of chemical physics} \textbf{2003},
  \emph{118}, 6188--6199\relax
\mciteBstWouldAddEndPuncttrue
\mciteSetBstMidEndSepPunct{\mcitedefaultmidpunct}
{\mcitedefaultendpunct}{\mcitedefaultseppunct}\relax
\EndOfBibitem
\bibitem[Honigmann \latin{et~al.}(2006)Honigmann, Buenker, and
  Liebermann]{honigmann2006complex}
Honigmann,~M.; Buenker,~R.~J.; Liebermann,~H.-P. Complex self-consistent field
  and multireference single-and double-excitation configuration interaction
  calculations for the $\Pi$g2 resonance state of N2-. \emph{The Journal of
  chemical physics} \textbf{2006}, \emph{125}\relax
\mciteBstWouldAddEndPuncttrue
\mciteSetBstMidEndSepPunct{\mcitedefaultmidpunct}
{\mcitedefaultendpunct}{\mcitedefaultseppunct}\relax
\EndOfBibitem
\bibitem[Honigmann \latin{et~al.}(2010)Honigmann, Liebermann, and
  Buenker]{honigmann2010use}
Honigmann,~M.; Liebermann,~H.-P.; Buenker,~R.~J. Use of complex configuration
  interaction calculations and the stationary principle for the description of
  metastable electronic states of HCl-. \emph{The Journal of chemical physics}
  \textbf{2010}, \emph{133}\relax
\mciteBstWouldAddEndPuncttrue
\mciteSetBstMidEndSepPunct{\mcitedefaultmidpunct}
{\mcitedefaultendpunct}{\mcitedefaultseppunct}\relax
\EndOfBibitem
\bibitem[Sommerfeld and Santra(2001)Sommerfeld, and
  Santra]{sommerfeld2001efficient}
Sommerfeld,~T.; Santra,~R. Efficient method to perform CAP/CI calculations for
  temporary anions. \emph{International Journal of Quantum Chemistry}
  \textbf{2001}, \emph{82}, 218--226\relax
\mciteBstWouldAddEndPuncttrue
\mciteSetBstMidEndSepPunct{\mcitedefaultmidpunct}
{\mcitedefaultendpunct}{\mcitedefaultseppunct}\relax
\EndOfBibitem
\bibitem[Sommerfeld \latin{et~al.}(1998)Sommerfeld, Riss, Meyer, Cederbaum,
  Engels, and Suter]{sommerfeld1998temporary}
Sommerfeld,~T.; Riss,~U.; Meyer,~H.; Cederbaum,~L.; Engels,~B.; Suter,~H.
  Temporary anions-calculation of energy and lifetime by absorbing potentials:
  The resonance. \emph{Journal of Physics B: Atomic, Molecular and Optical
  Physics} \textbf{1998}, \emph{31}, 4107\relax
\mciteBstWouldAddEndPuncttrue
\mciteSetBstMidEndSepPunct{\mcitedefaultmidpunct}
{\mcitedefaultendpunct}{\mcitedefaultseppunct}\relax
\EndOfBibitem
\bibitem[Jagau \latin{et~al.}(2014)Jagau, Zuev, Bravaya, Epifanovsky, and
  Krylov]{jagau2014fresh}
Jagau,~T.-C.; Zuev,~D.; Bravaya,~K.~B.; Epifanovsky,~E.; Krylov,~A.~I. A fresh
  look at resonances and complex absorbing potentials: Density matrix-based
  approach. \emph{The journal of physical chemistry letters} \textbf{2014},
  \emph{5}, 310--315\relax
\mciteBstWouldAddEndPuncttrue
\mciteSetBstMidEndSepPunct{\mcitedefaultmidpunct}
{\mcitedefaultendpunct}{\mcitedefaultseppunct}\relax
\EndOfBibitem
\bibitem[Belogolova \latin{et~al.}(2021)Belogolova, Dempwolff, Dreuw, and
  Trofimov]{belogolova2021complex}
Belogolova,~A.; Dempwolff,~A.; Dreuw,~A.; Trofimov,~A. \emph{J. Phys.: Conf.
  Ser.} \textbf{2021}, \emph{1847}, 012050\relax
\mciteBstWouldAddEndPuncttrue
\mciteSetBstMidEndSepPunct{\mcitedefaultmidpunct}
{\mcitedefaultendpunct}{\mcitedefaultseppunct}\relax
\EndOfBibitem
\bibitem[Dempwolff \latin{et~al.}(2021)Dempwolff, Belogolova, Sommerfeld,
  Trofimov, and Dreuw]{dempwolff2021cap}
Dempwolff,~A.~L.; Belogolova,~A.~M.; Sommerfeld,~T.; Trofimov,~A.~B.; Dreuw,~A.
  \emph{J. Chem. Phys.} \textbf{2021}, \emph{155}\relax
\mciteBstWouldAddEndPuncttrue
\mciteSetBstMidEndSepPunct{\mcitedefaultmidpunct}
{\mcitedefaultendpunct}{\mcitedefaultseppunct}\relax
\EndOfBibitem
\bibitem[Head-Gordon \latin{et~al.}(1995)Head-Gordon, Grana, Maurice, and
  White]{head1995analysis}
Head-Gordon,~M.; Grana,~A.~M.; Maurice,~D.; White,~C.~A. Analysis of electronic
  transitions as the difference of electron attachment and detachment
  densities. \emph{The Journal of Physical Chemistry} \textbf{1995}, \emph{99},
  14261--14270\relax
\mciteBstWouldAddEndPuncttrue
\mciteSetBstMidEndSepPunct{\mcitedefaultmidpunct}
{\mcitedefaultendpunct}{\mcitedefaultseppunct}\relax
\EndOfBibitem
\bibitem[Closser \latin{et~al.}(2014)Closser, Gessner, and
  Head-Gordon]{closser2014simulations}
Closser,~K.~D.; Gessner,~O.; Head-Gordon,~M. Simulations of the dissociation of
  small helium clusters with ab initio molecular dynamics in electronically
  excited states. \emph{The Journal of chemical physics} \textbf{2014},
  \emph{140}\relax
\mciteBstWouldAddEndPuncttrue
\mciteSetBstMidEndSepPunct{\mcitedefaultmidpunct}
{\mcitedefaultendpunct}{\mcitedefaultseppunct}\relax
\EndOfBibitem
\bibitem[Plasser \latin{et~al.}(2016)Plasser, Ruckenbauer, Mai, Oppel,
  Marquetand, and Gonz{\'a}lez]{plasser2016efficient}
Plasser,~F.; Ruckenbauer,~M.; Mai,~S.; Oppel,~M.; Marquetand,~P.;
  Gonz{\'a}lez,~L. Efficient and flexible computation of many-electron wave
  function overlaps. \emph{Journal of chemical theory and computation}
  \textbf{2016}, \emph{12}, 1207--1219\relax
\mciteBstWouldAddEndPuncttrue
\mciteSetBstMidEndSepPunct{\mcitedefaultmidpunct}
{\mcitedefaultendpunct}{\mcitedefaultseppunct}\relax
\EndOfBibitem
\bibitem[Aquilante \latin{et~al.}(2016)Aquilante, Autschbach, Carlson,
  Chibotaru, Delcey, De~Vico, Fdez.~Galván, Ferré, Frutos, Gagliardi,
  Garavelli, Giussani, Hoyer, Li~Manni, Lischka, Ma, Malmqvist, Müller, Nenov,
  Olivucci, Pedersen, Peng, Plasser, Pritchard, Reiher, Rivalta, Schapiro,
  Segarra-Martí, Stenrup, Truhlar, Ungur, Valentini, Vancoillie, Veryazov,
  Vysotskiy, Weingart, Zapata, and Lindh]{MOLCAS8}
Aquilante,~F. \latin{et~al.}  Molcas 8: New capabilities for
  multiconfigurational quantum chemical calculations across the periodic table.
  \emph{Journal of Computational Chemistry} \textbf{2016}, \emph{37},
  506--541\relax
\mciteBstWouldAddEndPuncttrue
\mciteSetBstMidEndSepPunct{\mcitedefaultmidpunct}
{\mcitedefaultendpunct}{\mcitedefaultseppunct}\relax
\EndOfBibitem
\bibitem[Wang and Song(2016)Wang, and Song]{wang2016geometry}
Wang,~L.-P.; Song,~C. Geometry optimization made simple with translation and
  rotation coordinates. \emph{The Journal of chemical physics} \textbf{2016},
  \emph{144}\relax
\mciteBstWouldAddEndPuncttrue
\mciteSetBstMidEndSepPunct{\mcitedefaultmidpunct}
{\mcitedefaultendpunct}{\mcitedefaultseppunct}\relax
\EndOfBibitem
\bibitem[Heide and King(2020)Heide, and King]{heide2020optking}
Heide,~A.; King,~R. OPTKING: A Python version of the PSI4 geometry optimizer.
  \emph{For the current version, see https://github. com/psi-rking/optking}
  \textbf{2020}, \relax
\mciteBstWouldAddEndPunctfalse
\mciteSetBstMidEndSepPunct{\mcitedefaultmidpunct}
{}{\mcitedefaultseppunct}\relax
\EndOfBibitem
\bibitem[Gayvert()]{gayvert}
Gayvert,~J.~R. OpenCAP: An open-source program for studying resonances in
  molecules. \url{https://github.com/gayverjr/opencap}\relax
\mciteBstWouldAddEndPuncttrue
\mciteSetBstMidEndSepPunct{\mcitedefaultmidpunct}
{\mcitedefaultendpunct}{\mcitedefaultseppunct}\relax
\EndOfBibitem
\bibitem[Lischka \latin{et~al.}(2011)Lischka, Müller, Szalay, Shavitt, Pitzer,
  and Shepard]{COLUMBUS}
Lischka,~H.; Müller,~T.; Szalay,~P.~G.; Shavitt,~I.; Pitzer,~R.~M.;
  Shepard,~R. Columbus—a program system for advanced multireference theory
  calculations. \emph{WIREs Computational Molecular Science} \textbf{2011},
  \emph{1}, 191--199\relax
\mciteBstWouldAddEndPuncttrue
\mciteSetBstMidEndSepPunct{\mcitedefaultmidpunct}
{\mcitedefaultendpunct}{\mcitedefaultseppunct}\relax
\EndOfBibitem
\bibitem[Szalay \latin{et~al.}(2012)Szalay, Muller, Gidofalvi, Lischka, and
  Shepard]{szalay2012multiconfiguration}
Szalay,~P.~G.; Muller,~T.; Gidofalvi,~G.; Lischka,~H.; Shepard,~R.
  Multiconfiguration self-consistent field and multireference configuration
  interaction methods and applications. \emph{Chemical reviews} \textbf{2012},
  \emph{112}, 108--181\relax
\mciteBstWouldAddEndPuncttrue
\mciteSetBstMidEndSepPunct{\mcitedefaultmidpunct}
{\mcitedefaultendpunct}{\mcitedefaultseppunct}\relax
\EndOfBibitem
\bibitem[Lischka \latin{et~al.}(2004)Lischka, Dallos, Szalay, Yarkony, and
  Shepard]{lischka2004analytic}
Lischka,~H.; Dallos,~M.; Szalay,~P.~G.; Yarkony,~D.~R.; Shepard,~R. Analytic
  evaluation of nonadiabatic coupling terms at the MR-CI level. I. Formalism.
  \emph{The Journal of chemical physics} \textbf{2004}, \emph{120},
  7322--7329\relax
\mciteBstWouldAddEndPuncttrue
\mciteSetBstMidEndSepPunct{\mcitedefaultmidpunct}
{\mcitedefaultendpunct}{\mcitedefaultseppunct}\relax
\EndOfBibitem
\bibitem[Dallos \latin{et~al.}(2004)Dallos, Lischka, Shepard, Yarkony, and
  Szalay]{dallos2004analytic}
Dallos,~M.; Lischka,~H.; Shepard,~R.; Yarkony,~D.~R.; Szalay,~P.~G. Analytic
  evaluation of nonadiabatic coupling terms at the MR-CI level. II. Minima on
  the crossing seam: Formaldehyde and the photodimerization of ethylene.
  \emph{The Journal of chemical physics} \textbf{2004}, \emph{120},
  7330--7339\relax
\mciteBstWouldAddEndPuncttrue
\mciteSetBstMidEndSepPunct{\mcitedefaultmidpunct}
{\mcitedefaultendpunct}{\mcitedefaultseppunct}\relax
\EndOfBibitem
\bibitem[Berman \latin{et~al.}(1983)Berman, Estrada, Cederbaum, and
  Domcke]{berman1983nuclear}
Berman,~M.; Estrada,~H.; Cederbaum,~L.~S.; Domcke,~W. Nuclear dynamics in
  resonant electron-molecule scattering beyond the local approximation: The
  2.3-eV shape resonance in N 2. \emph{Physical Review A} \textbf{1983},
  \emph{28}, 1363\relax
\mciteBstWouldAddEndPuncttrue
\mciteSetBstMidEndSepPunct{\mcitedefaultmidpunct}
{\mcitedefaultendpunct}{\mcitedefaultseppunct}\relax
\EndOfBibitem
\bibitem[Choi and Jordan(1989)Choi, and Jordan]{choi1989electron}
Choi,~Y.; Jordan,~K. Electron transmission spectra of carbonyl fluoride:
  Determination of the vertical electron affinity. \emph{Chemical physics
  letters} \textbf{1989}, \emph{156}, 450--454\relax
\mciteBstWouldAddEndPuncttrue
\mciteSetBstMidEndSepPunct{\mcitedefaultmidpunct}
{\mcitedefaultendpunct}{\mcitedefaultseppunct}\relax
\EndOfBibitem
\bibitem[Kunitsa and Bravaya(2019)Kunitsa, and Bravaya]{kunitsa2019feshbach}
Kunitsa,~A.~A.; Bravaya,~K.~B. Feshbach projection XMCQDPT2 model for
  metastable electronic states. \emph{arXiv preprint arXiv:1906.11390}
  \textbf{2019}, \relax
\mciteBstWouldAddEndPunctfalse
\mciteSetBstMidEndSepPunct{\mcitedefaultmidpunct}
{}{\mcitedefaultseppunct}\relax
\EndOfBibitem
\bibitem[Burrow and Michejda(1976)Burrow, and Michejda]{burrow1976electron}
Burrow,~P.; Michejda,~J. Electron transmission study of the formaldehyde
  electron affinity. \emph{Chemical Physics Letters} \textbf{1976}, \emph{42},
  223--226\relax
\mciteBstWouldAddEndPuncttrue
\mciteSetBstMidEndSepPunct{\mcitedefaultmidpunct}
{\mcitedefaultendpunct}{\mcitedefaultseppunct}\relax
\EndOfBibitem
\bibitem[Benda \latin{et~al.}(2018)Benda, Rickmeyer, and
  Jagau]{benda2018structure}
Benda,~Z.; Rickmeyer,~K.; Jagau,~T.-C. Structure optimization of temporary
  anions. \emph{Journal of Chemical Theory and Computation} \textbf{2018},
  \emph{14}, 3468--3478\relax
\mciteBstWouldAddEndPuncttrue
\mciteSetBstMidEndSepPunct{\mcitedefaultmidpunct}
{\mcitedefaultendpunct}{\mcitedefaultseppunct}\relax
\EndOfBibitem
\end{mcitethebibliography}


\begin{thebibliography}{99}
	
	\bibitem{benda2017capeom}
	Zsuzsanna Benda and Thomas-C. Jagau, 
	\newblock Communication: Analytic gradients for the complex absorbing potential equation-of-motion coupled-cluster method,
	\newblock \textit{The Journal of Chemical Physics}, \textbf{146}, 3 (2017).
	
	\bibitem{gayvert2022projectedeom}
	James R. Gayvert and Ksenia B. Bravaya, 
	\newblock Projected CAP-EOM-CCSD method for electronic resonances,
	\newblock \textit{The Journal of Chemical Physics}, \textbf{156}, 9 (2022).
	
	\bibitem{head1995analysisADden}
	Martin Head-Gordon, Ana M. Grana, David Maurice, and Christopher A. White,
	\newblock Analysis of electronic transitions as the difference of electron attachment and detachment densities,
	\newblock \textit{The Journal of Physical Chemistry}, \textbf{99}, 39, 14261--14270 (1995).
	
	\bibitem{closser2014simulationstrack}
	Kristina D. Closser, Oliver Gessner, and Martin Head-Gordon, 
	\newblock Simulations of the dissociation of small helium clusters with ab initio molecular dynamics in electronically excited states,
	\newblock \textit{The Journal of Chemical Physics}, \textbf{140}, 13 (2014).
	
	\bibitem{qchem}
	Evgeny Epifanovsky, Andrew TB Gilbert, Xintian Feng, Joonho Lee, Yuezhi Mao, Narbe Mardirossian, Pavel Pokhilko, Alec F White, Marc P Coons, Adrian L Dempwolff, \textit{et al.}, 
	\newblock Software for the frontiers of quantum chemistry: An overview of developments in the Q-Chem 5 package,
	\newblock \textit{The Journal of Chemical Physics}, \textbf{155}, 084801 (2021).
	
	
\end{thebibliography}
